\documentclass{aa}  

\usepackage{graphicx}
\usepackage{subcaption}
\usepackage{amsmath}
\usepackage{txfonts}
\usepackage{hyperref}
\usepackage{subcaption}
\usepackage[normalem]{ulem}
\usepackage[x11names]{xcolor}
\usepackage{academicons}
\usepackage[flushleft]{threeparttable}
\usepackage{booktabs,caption}
\usepackage{xcolor}
\usepackage{soul}

\begin{document}

   \title{Role of magnetic shear distribution on the formation of eruptive flux ropes}

   \author{Samrat Sen$^*$ \inst{1, 2} \href{https://orcid.org/0000-0003-1546-381X}{\includegraphics[scale=0.05]{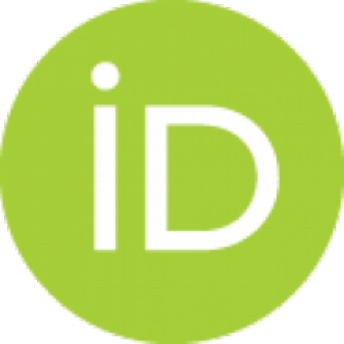}}, Sushree S Nayak \inst{3, 4, 5, 6} \href{https://orcid.org/0000-0002-4241-627X}{\includegraphics[scale=0.05]{figures/orcid-ID.pdf}}, Patrick Antolin \inst{7} \href{https://https://orcid.org/0000-0003-1529-4681}{\includegraphics[scale=0.05]{figures/orcid-ID.pdf}}
   }

   \institute{
   Instituto de Astrof\'{i}sica de Canarias, 38205 La Laguna, Tenerife, Spain
   \and
   Universidad de La Laguna, 38206 La Laguna, Tenerife, Spain 
   \and 
   Departament de F\'isica, Universitat de les Illes Balears, E-07122,
    Palma de Mallorca, Spain
      \and
      Institute of Applied Computing and Community
    Code (IAC3), UIB, E-07122, Palma de Mallorca, Spain
    \and 
    Indian Institute of Astrophysics, Koramangala, Bangalore 560034, India
    \and
    Center for Space Plasma and Aeronomic Research, The University of Alabama in Huntsville, Huntsville, AL 35899, USA
    \and
    School of Engineering, Physics, and Mathematics, Northumbria University, Newcastle upon Tyne, NE1 8ST, UK
    \\
   $^*$\email{samratseniitmadras@gmail.com; samrat.sen@iac.es}
         }
         
   \date{Received: XXXX; accepted: XXXX}
   \date{}
 
  \abstract
   {Erupting flux ropes play crucial role in powering a wide range of solar transients, including flares, jets, and coronal mass ejections. These events are driven by the release of stored magnetic energy, facilitated by the shear in the complex magnetic topologies. However, the mechanisms governing the formation and eruption of flux ropes, particularly the role of magnetic shear distribution in coronal arcades are not fully understood.}
   {We investigate how the spatial distribution of magnetic shear along coronal arcades influences the formation and evolution of eruptive flux ropes, with a focus on the evolution of mean shear during different phases of the eruption process.}
   {We employ 2.5D resistive magnetohydrodynamic (MHD) simulations incorporating nonadiabatic effects of optically thin radiative losses, magnetic field-aligned thermal conduction, and spatially varying background heating, to realistically model the coronal environment. A stratified solar atmosphere under gravity is initialized with a non-force-free field comprising sheared arcades. We study two different cases by varying the initial shear to analyze their resulting dynamics, and the possibility of flux rope formation and eruptions.}   
   {Our results show that strong initial magnetic shear leads to spontaneous flux rope formation and eruption via magnetic reconnection, driven by Lorentz force. The persistence and distribution of shear along the arcades are crucial in determining the formation and onset of flux rope instabilities. The shear distribution infers the non-potentiality distributed along arcades and demonstrates its relevance in identifying sites prone to eruptive activity. The evolution of mean shear and the relative strength between guide to reconnection fields during the pre- and post-eruption phases are explored, with implications of bulk heating for the ``hot onset'' phenomena in flares, and particle acceleration. On the other hand, the weaker shear case does not lead to formation of any flux ropes.}
   {The spatial distribution of magnetic shear and its evolution, and mean shear play a decisive role in the dynamics of flux rope formation and eruption. Our findings highlight the limitations of relying solely on foot point shear and underscore the need for coronal scale diagnostics. These results are relevant for understanding eruptive onset conditions and can promote a better interpretation of coronal observations from current and future missions.}
 
   \keywords{instabilities -- Magnetic reconnection -- Magnetohydrodynamics (MHD) -- Methods: numerical -- Sun: corona}

\titlerunning{Magnetic shear distribution for eruptive flux ropes}
\authorrunning{Sen et al.}

\maketitle

\section{Introduction} \label{S-Introduction} 

Magnetic reconnection is widely accepted as one of the fundamental processes in powering many solar transients like solar flares, jets, coronal mass ejections (CMEs) etc. It involves release of stressed magnetic energy with conversion into heat, kinetic energy, change of magnetic topology and accelerates particles from the reconnection sites. Key to most of these reconnection driven transients are the mechanisms of energy build-up, storage and release which are still poorly understood.  

Complex magnetic topology plays an important role in triggering these energetics. The magnetic energy build-up process rely on the stressing/stretching of the magnetic loops anchored at the photosphere. Owing to the photospheric motions, the field lines develop shear in them. Basically, these constant motions on the photosphere provide non-potentiality to the flux tubes which help in storing the energy in the system. Important factors contributing to the non-potentiality are shearing in the loops with respect to the polarity inversion line (PIL) \citep{1996ApJ...467..881S} and the magnetic complexity involved in the flux distribution like multi-polar sunspots having different positioning of umbra and penumbra \citep{1986SoPh..103..111P, 1998ApJ...502L.181A}. 

Seminal works by \citet{1984SoPh...91..115H, 1986AdSpR...6f...7H} have shown the role of angular shear in the regions hosting flares and provided a threshold  of a minimum of $80 {^\circ}$ of shearing angle for the flare producing regions. In their revised definition of shearing angle, \citet{1993SoPh..148..119L} calculated the angle between the observed magnetic field and its corresponding current-free field. They show the correlation of the high shearing angle to the flare onset from the vector magnetograph measurements. In the work by \cite{1992SoPh..138..353S}, they highlighted the change in shear angle as the deciding factor for the onset of a flare. \citet{2023ApJ...955...34Q} have shown an intermediate angle of $\leq 40 {^\circ}$, needed for particle acceleration from measurements of the shear in post-reconnection flare loops. With the constant change of boundary, if we consider photosphere per se, the non-potentiality in the active region develops  due to the change in flux emergence \citep{1992PASJ...44L.173S,2007Sci...318.1591S,2008ApJ...673L.211M,2021NatCo..12.6621M} and/or cancellations \citep{2008A&A...481L..57C,2011ApJ...738L..20H,2014PASJ...66S..12Y,2016ApJ...822L..23P,2018ApJ...853..189P,2017ApJ...844...28S,2019ApJ...882...16M}, and rotational motions of the sunspots \citep{Magara:2009}. How the temporal evolutions of non-potentiality in the active region due to bottom boundary changes affect the reconnection driven transients or eruptions is still a not-so clearly understood problem. 

However, magnetic shear can persist in the solar corona even after the cessation of photospheric motions that initially generated it, and allows the magnetic structures to maintain their configuration over extended periods. Studies have shown that active regions can retain significant magnetic helicity for a sufficiently long duration (several months), indicating that shear can remain long after the initial photospheric activity has subsided \citep{Demoulin:2009}. Also, the magnetic structure in solar corona can undergo reconfiguration due to the transfer of magnetic helicity, leading to the development of shear in coronal loops \citep{Yeates:2024}. These scenarios illustrate that magnetic shear in coronal arcades does not always necessitate concurrent photospheric activity. Solar corona allows for the persistence and evolution of shear through various mechanisms as stated above. The formation, stability, and evolution of coronal flux ropes depend on the magnetic shear (and twist) distributed along the arcades, rather than the shear solely at their foot points. Although shear at the foot points may appear strong due to localized photospheric motions, it does not necessarily contribute to flux rope formation into the overlying arcades. Instabilities that develop in flux ropes, such as those arising from magnetic reconnection \citep{cheng:2001,cheng:2003}, torus instability \citep{2006PhRvL..96y5002K}, or kink instability \citep{2004A&A...413L..27T}, are governed by the overall magnetic configuration and not by the foot point shear alone. In fact, the spatial distribution of shear helps identify regions in the arcade where the instability can trigger, which is essential for predicting eruptions. 

The evolution of twist along the magnetic field lines at the pre- and post flare stages is reported from the extrapolated magnetic field lines for an active region \citep{Inoue:2011}. In \cite{Aulanier:2012}, the dispersion of shear angle along the polarity inversion line (PIL) as a function of the loop apex height for the post-flare loops is reported in a magnetohydrodynamic (MHD) simulation. Also, the study shows the evolution of the shear angle at a few selected points on the loop for the pre- and post-flare stages. However, this model is restricted for a zero-$\beta$ case, and the study does not give much attention on the shear distribution along the entire field lines. Therefore, it warrants a deeper investigation of the shear distribution along magnetic arcades in the formation of eruptive flux ropes, as well as the evolution of mean shear during the flux rope build-up and post-eruption phases. To realize this aspect, we define mean shear as the shear per unit length along a magnetic arcade, which is a more relevant quantity than foot point shear for representing the overall non-potentiality and magnetic stress distributed throughout the coronal structure. We use a series of 2.5D resistive MHD simulation reported in \citep[][S24 hereafter]{Sen:2024} by varying the initial shear of the arcades, and using a nonadiabatic framework that incorporates optically thin radiative losses, (field-aligned) thermal conduction, and steady, spatially varying background heating. Incorporation of the nonadiabatic terms in our simulation makes it more realistic (than adiabatic environment) for coronal medium. To replicate coronal loop structures, we employ a bipolar magnetic field configuration consisting of sheared arcades that are non-forcefree and embedded within a stratified solar atmosphere under the influence of gravity. The shear in the magnetic loops are associated with the guide field component. For the simulation with a high shear case, the system's Lorentz force, present from the initial state, drives a rapid evolution wherein flux ropes develop from the original bipolar field lines and ascend continuously toward eruption. This process is facilitated by spontaneous magnetic reconnections, a common mechanism in such scenarios. We also comment on the variability of the guide field component, which can be a deciding factor in the efficiency of bulk heating and particle acceleration during different phases of the eruptions.    

The paper is organized as follows. Section \ref{sec:setup} provides a brief description of the model, its initial configuration, and the boundary condition used. In Section \ref{sec:results}, we present the results along with a detailed analysis of the underlying physics. Section \ref{sec:summary} summarizes the key findings, connects them with existing and upcoming observations, and highlights the novelty of the work. Finally, in the conclusion, we discuss the caveats of the model and outline the potential directions for future improvements.   

\section{Model description} \label{sec:setup}
To investigate the role of magnetic shear in the formation of eruptive flux ropes, we used a 2.5D resistive-MHD simulation using MPI-parallelized adaptive mesh refinement versatile advection code (MPI-AMRVAC) \footnote{Open source at: \href{http://amrvac.org}{http://amrvac.org/}} \citep{2012JCoPh.231..718K, 2014ApJS..214....4P, 2018ApJS..234...30X, keppens2021, keppens2023}. 

\subsection{Initial condition}
The simulation domain spanned a horizontal range of $x=0-2\pi$ Mm, and a vertical range of $y=0-25$ Mm. The choice of the length scales in our model is motivated to address the miniature coronal loops, those can have the length $\sim 1$ Mm as observed in \cite{2013:Peter}. To emulate the magnetic topology of sheared magnetic loops, the initial magnetic field configuration of the system was specified as follows:    

\begin{align}
    B_x &= B_0 \ \text{sin}(k_x(x-x_0)) \ \text{exp}(-yk_x/\sigma)\,, \label{eq:bx}\\ \label{eq:by}
    B_y &= B_0 \sigma \ \text{cos}(k_x(x-x_0)) \ \text{exp}(-yk_x/\sigma)\,,\\ \label{eq:bz}
    B_z &= B_0 \ \text{tan}(\gamma) \ \text{sin}(k_x(x-x_0)) \ \text{exp}(-yk_x/\sigma) \,,
\end{align}
where, $B_0=9$ G, $\sigma=10$ is a dimensionless parameter,  $k_x = 2\pi/L_x$ ($L_x=2\pi$ Mm is the horizontal span of the simulation domain), which gives $k_x=1$ Mm$^{-1}$. We used the initial shear angle, $\gamma = 72.5^\circ,$ and $25.8^\circ$ for two different case studies in the simulation keeping all the other conditions same. The shear angle (or, shear density) of the projected field lines at the $y = 0$ plane with the $+x$ axis is,
\begin{align}\label{eq:shear_angle}
    \gamma = \mathrm{arctan}\bigg(\frac{B_z}{B_x}\bigg).
\end{align} 

The plasma density in the vertical direction is stratified according to hydrostatic equilibrium under gravity, where ${\bf g} = -g(y) \mathbf{e_y}$ and
\begin{align}
g(y) = g_0 \frac{R_{\odot}^2}{(R_{\odot}+y)^2},
\end{align}
with $g_0 = 274$ m s$^{-2}$ representing the gravitational acceleration at the solar surface, and $R_{\odot} = 695.7$ Mm denoting the solar radius. An isothermal atmosphere with temperature $T_0 = 1$ MK is assumed as the initial condition. This isothermal approximation is appropriate given that the vertical extent of the domain is approximately 25 Mm. The corresponding initial density profile is given by
\begin{align}\label{eq:HS_equlb}
\rho_i(y) = \rho_0\ \text{exp}\left[-y/H(y)\right],
\end{align}
where $\displaystyle{H(y) = \frac{\mathcal{R} T_0}{\mu g_0} \frac{R_{\odot} +y}{R_{\odot}}}$ denotes the density scale height, determined using the gas constant $\mathcal{R}$ and the mean particle mass $\mu$. To reflect typical coronal conditions, the base density is set to $\rho_0 = 3.2 \times 10^{-15}$ g cm$^{-3}$. The model incorporates Spitzer-type thermal conduction, which is strictly field-aligned, with a conductivity given by ${\bf \kappa}_{||} = 10^{-6}, T^{5/2}$ erg cm$^{-1}$ s$^{-1}$ K$^{-1}$. Additionally, a steady background heating is included to precisely balance the initial optically thin radiative losses. The readers are referred to S24 for more details of the governing equations and numerical schemes used in the simulations. Note that the choice of the parameter, $\sigma$ is set in such a way that it corresponds to a non-force-free-field configuration at the initial state ($t=0$). The choice is motivated to avoid the computational time needed to build an unstable magnetic configuration from the initial mechanical equilibrium state, similar to attempts made with MHD simulations initiated with non-force-free-field setups as reported by \citet{2016:sanjay, 2024:Nayak}, and references therein. The system subsequently achieves a semi-equilibrium state (mechanical and thermal) at around $t=5.44$ min, and remains till $t=7$ min, where the evolution of the system is quasi-static in this time window compared to the earlier phase of the evolution (S24). According to the definition in Eq. \ref{eq:shear_angle}, the shear angle, $|\gamma|$ can vary between $0$ to $90^\circ$. These angles correspond to no shear ($\gamma=0$), when the a magnetic loop is perpendicular to the PIL, and to a maximum shear ($|\gamma|=90^\circ$) when they are parallel to each other. As already mentioned previously, we conducted simulations for two distinct cases: with $\gamma = 72.5^\circ$, and $25.8^\circ$ (keeping all the other parameters same), which we refer as ``simulation 1'' and ``simulation 2'', respectively. These two cases represent significantly different shear conditions, with the high-shear, and weak-shear cases, respectively.

\subsection{Boundary condition}
Periodic boundary conditions were applied at the lateral boundaries. At the top and bottom boundaries, magnetic fields were extrapolated using third-order and second-order zero-gradient methods, respectively. We also used the divergence of \textbf{B} cleaning with a parabolic diffusion method in the whole domain \citep{keppens2003, keppens2021}. Pressure and density at the bottom boundary were fixed to their corresponding local initial values. We copied the instantaneous temperature values of the top edge cells to the corresponding ghost cells at the top boundary, and then the density and pressure were specified based on the hydrostatic assumption, as described in \cite{zhao2017}. The velocity components at the bottom boundary were imposed antisymmetrically via mirror reflection to populate the ghost cells. At the top boundary, a third-order extrapolation with a zero-gradient condition was employed for all velocity components, with the additional constraint of no inflow.     

\begin{figure*}[hbt!]
    \centering
    \begin{subfigure}{0.6\columnwidth}
        \includegraphics[width=1\linewidth]{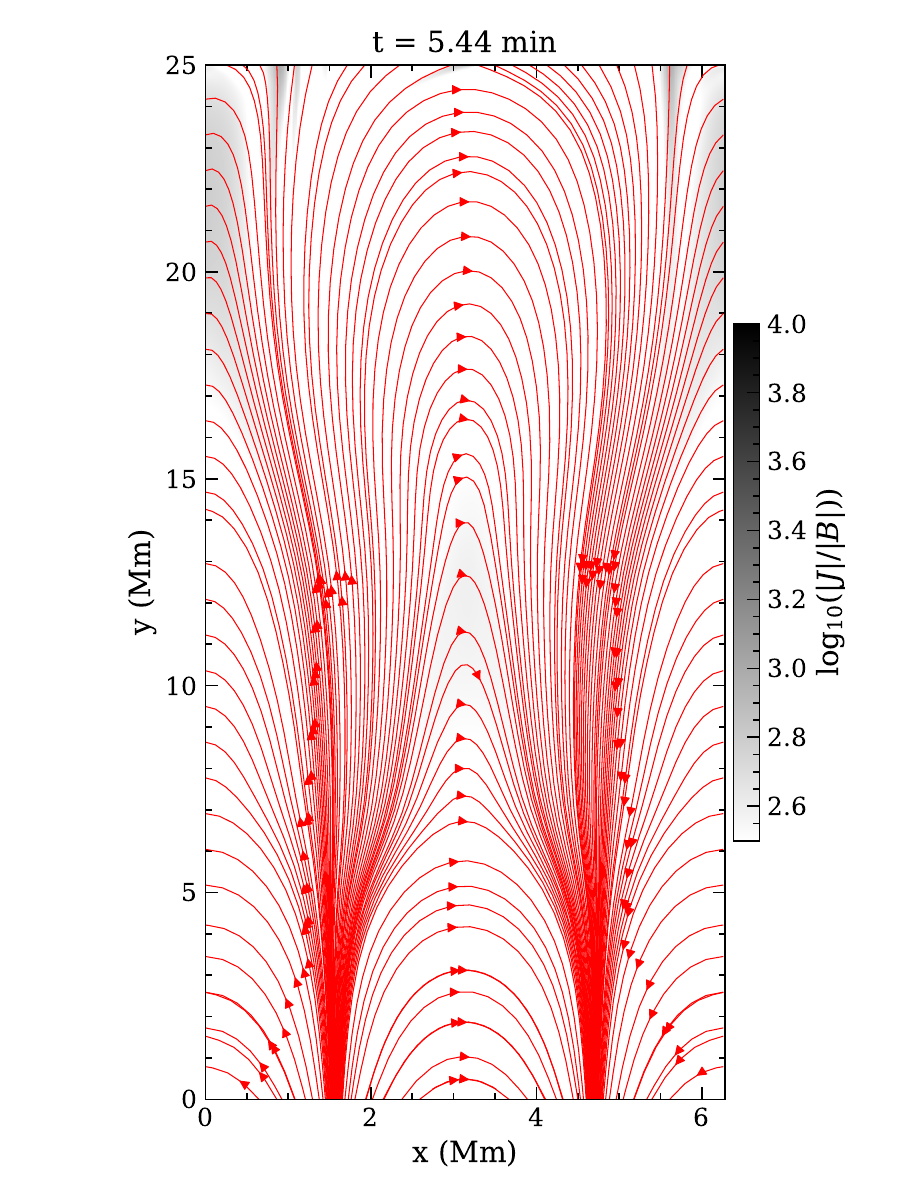}
        \caption{}
        \label{fig1_tl}
    \end{subfigure}
    \hspace{-0.1cm}
   \begin{subfigure}{0.6\columnwidth}
       \includegraphics[width=1\linewidth]{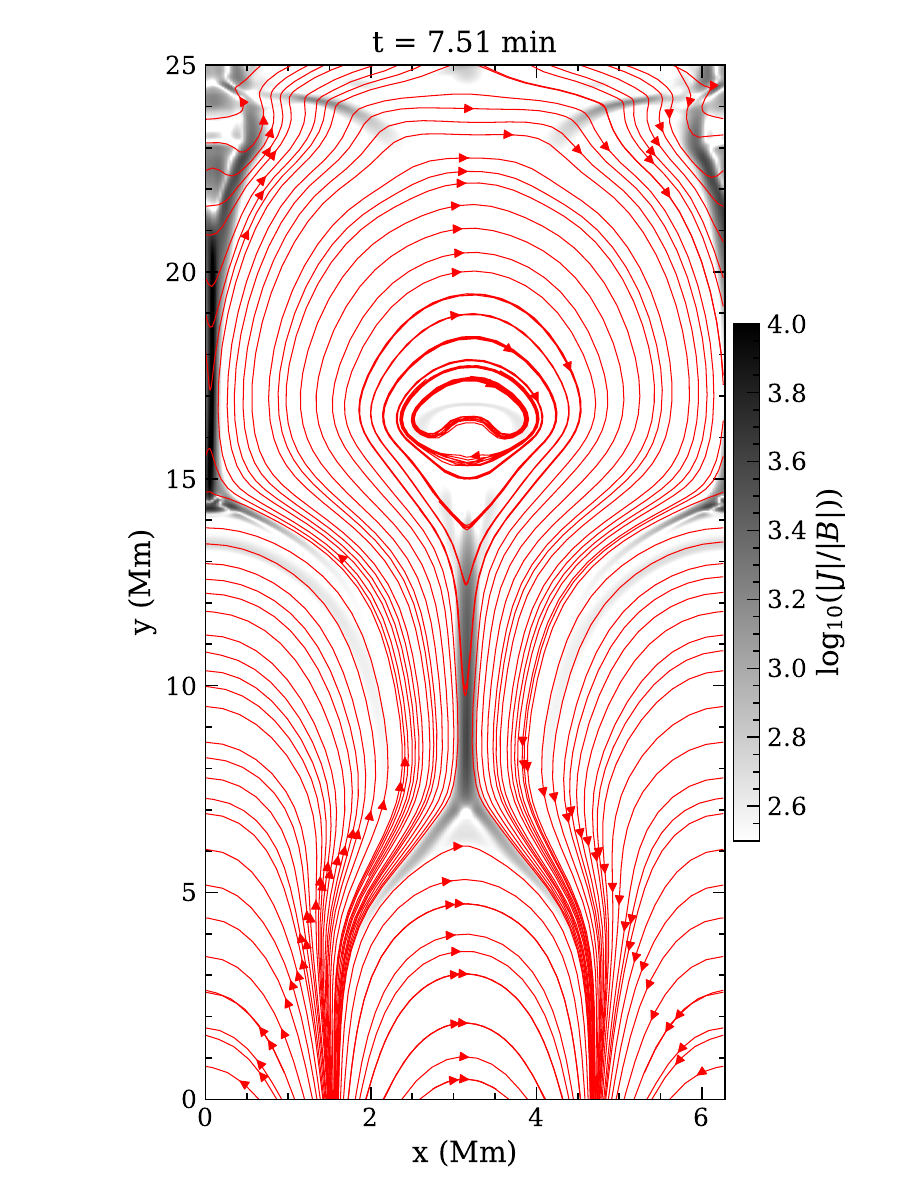}
       \caption{}
       \label{fig1_tm}
   \end{subfigure}
   \hspace{-0.1cm}
   \begin{subfigure}{0.6\columnwidth}
       \includegraphics[width=1\linewidth]{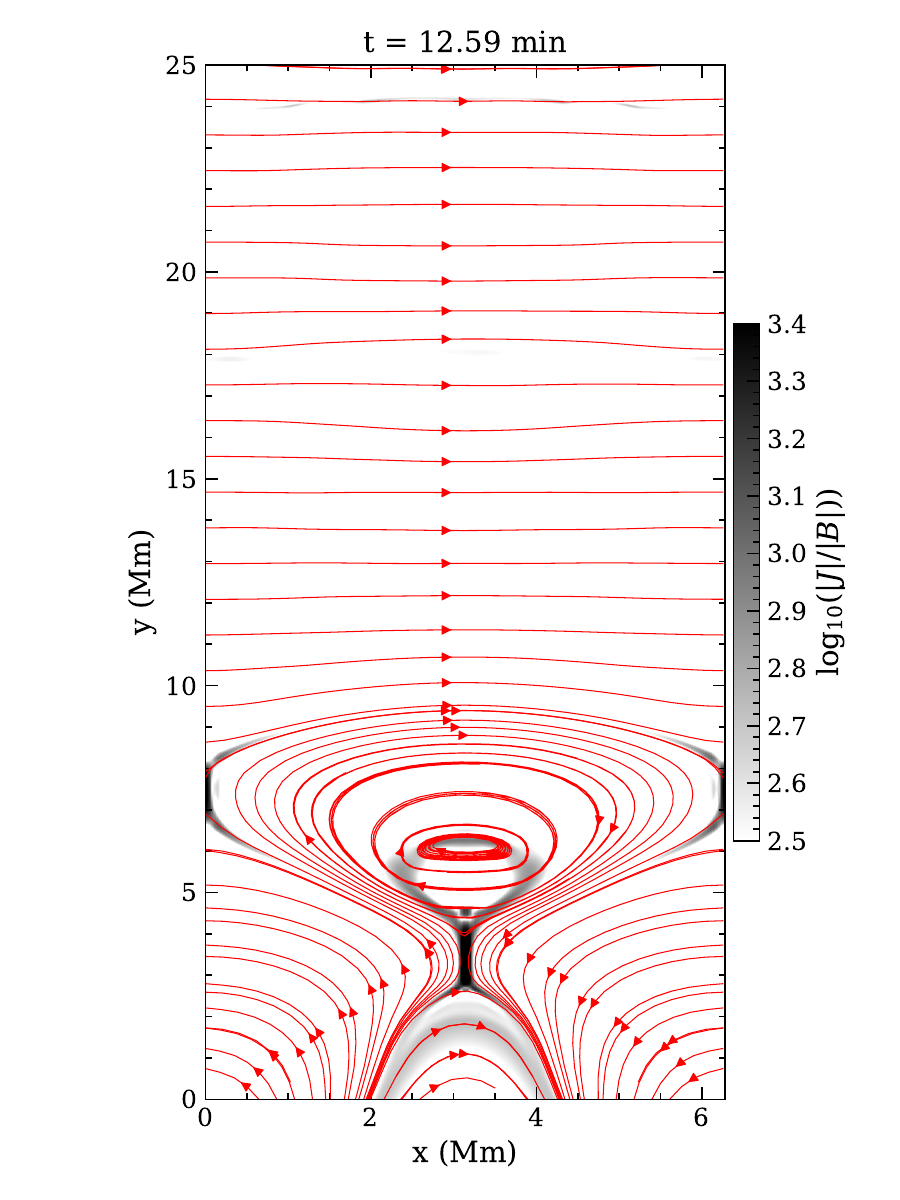}
       \caption{}
       \label{fig1_tr}
   \end{subfigure}

   \begin{subfigure}{0.6\columnwidth}
       \includegraphics[width=1\linewidth]{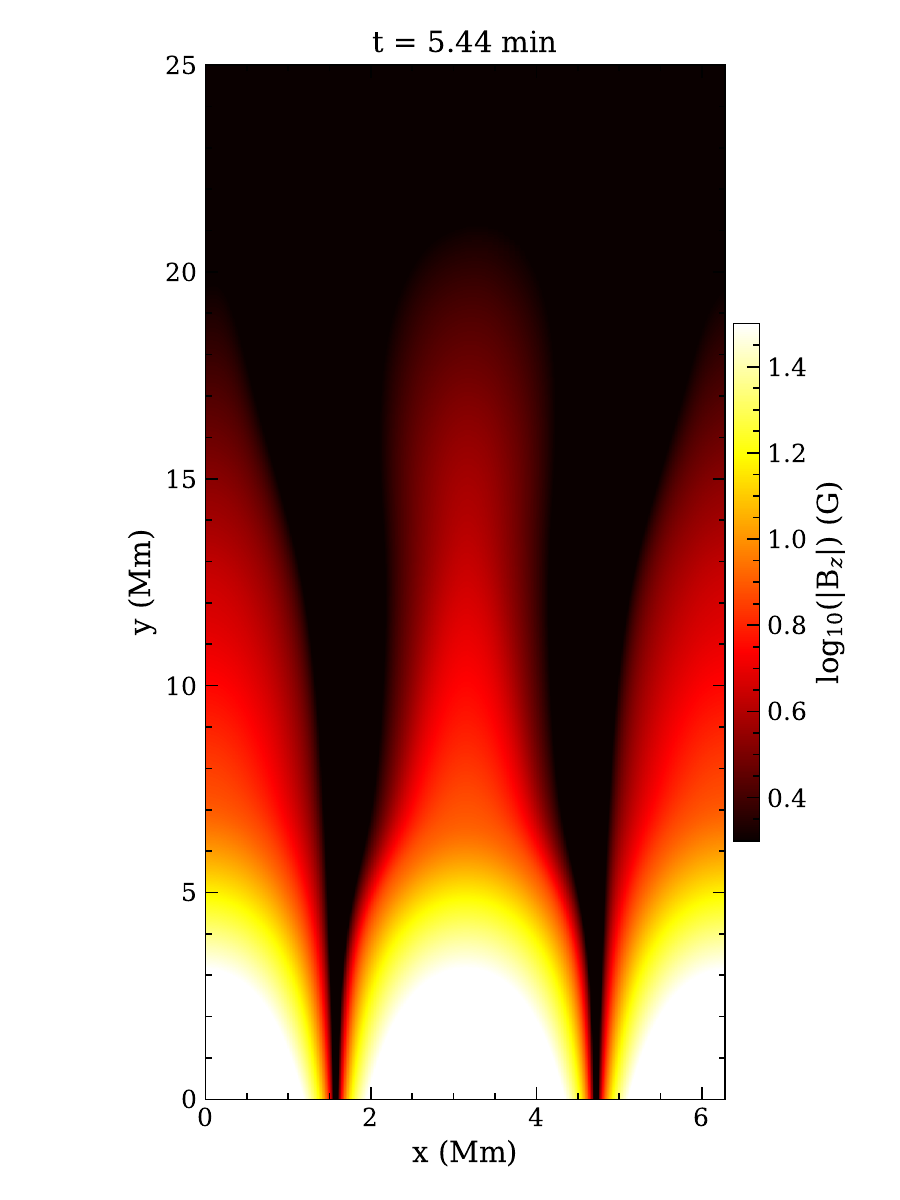}  
       \caption{}
       \label{fig1_bl}
   \end{subfigure}
   \hspace{-0.1cm}
    \begin{subfigure}{0.6\columnwidth}
        \includegraphics[width=1\linewidth]{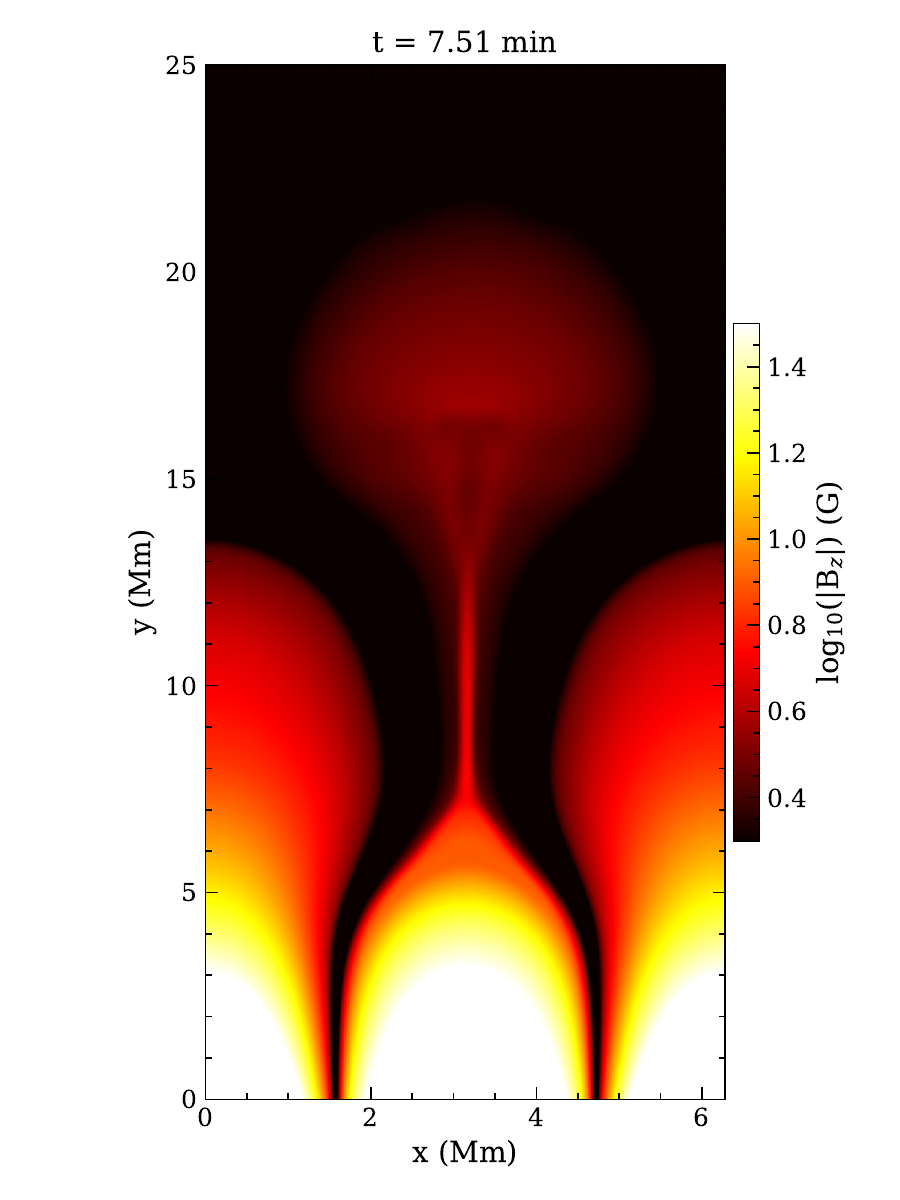}
        \caption{}
        \label{fig1_bm}
    \end{subfigure}
    \hspace{-0.1cm}
   \begin{subfigure}{0.6\columnwidth}
       \includegraphics[width=1\linewidth]{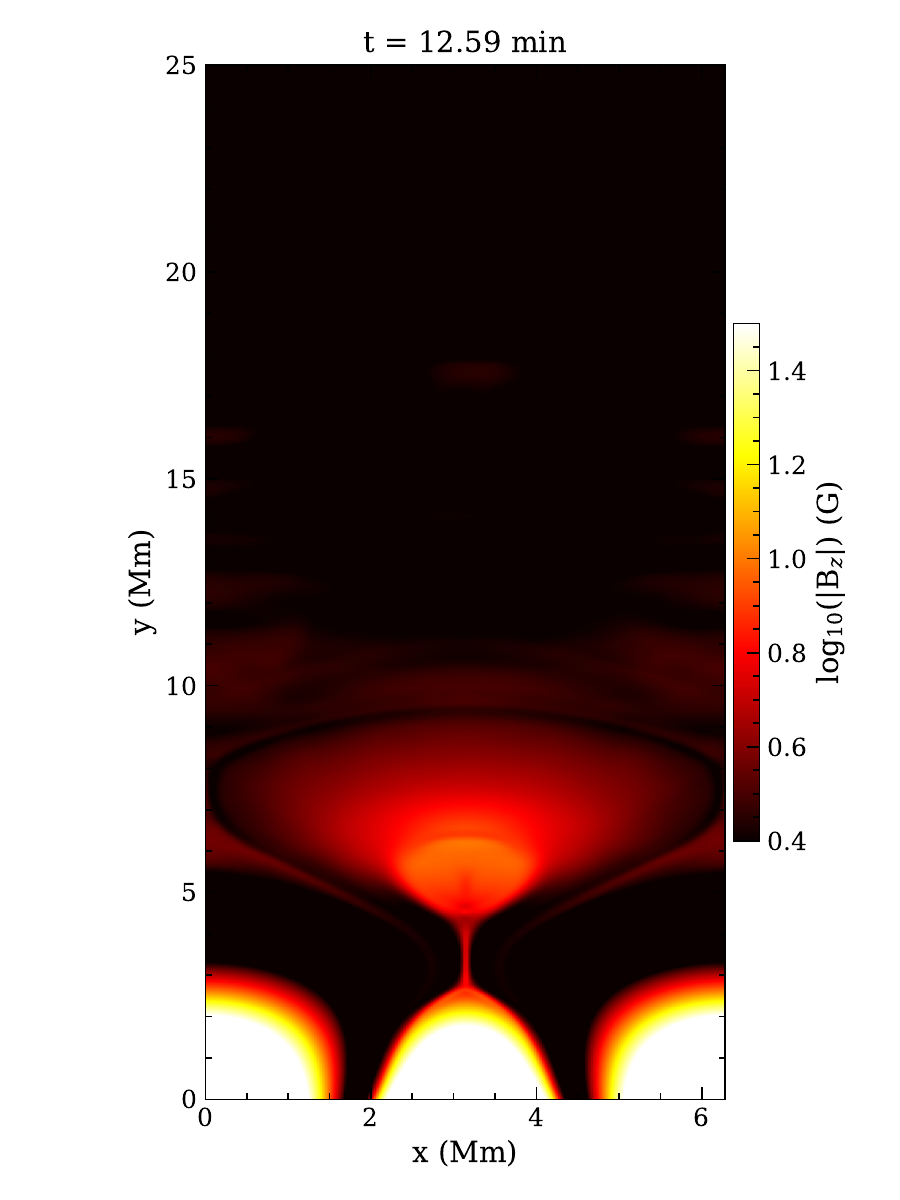}
       \caption{}
       \label{fig1_br}
   \end{subfigure}
 \caption{\textit{Top row:} Spatial distribution of $|\bf{J}|/|B|$ (in arbitrary units in log scale), and the projected magnetic field lines (in red) in the $x-y$ plane for (a) $t=5.44$, (b) $t=7.51$, and (c) $12.59$ s obtained from ``simulation 1''. The saturation level of the color bars are chosen appropriately as shown in the legends for a better visualization of the CS structures. \textit{Bottom row:} (d), (e), and (f) are same as the corresponding top panels, for the $|B_z|$ in logarithm scale, where the saturation level of the color bars are chosen between as shown in the respective legends for a better visualization.}
    \label{fig:cs_FL_pmag}
\end{figure*} 

\section{Results and analysis} \label{sec:results} 

\begin{figure*}
    \centering
\includegraphics[width=0.7\linewidth]{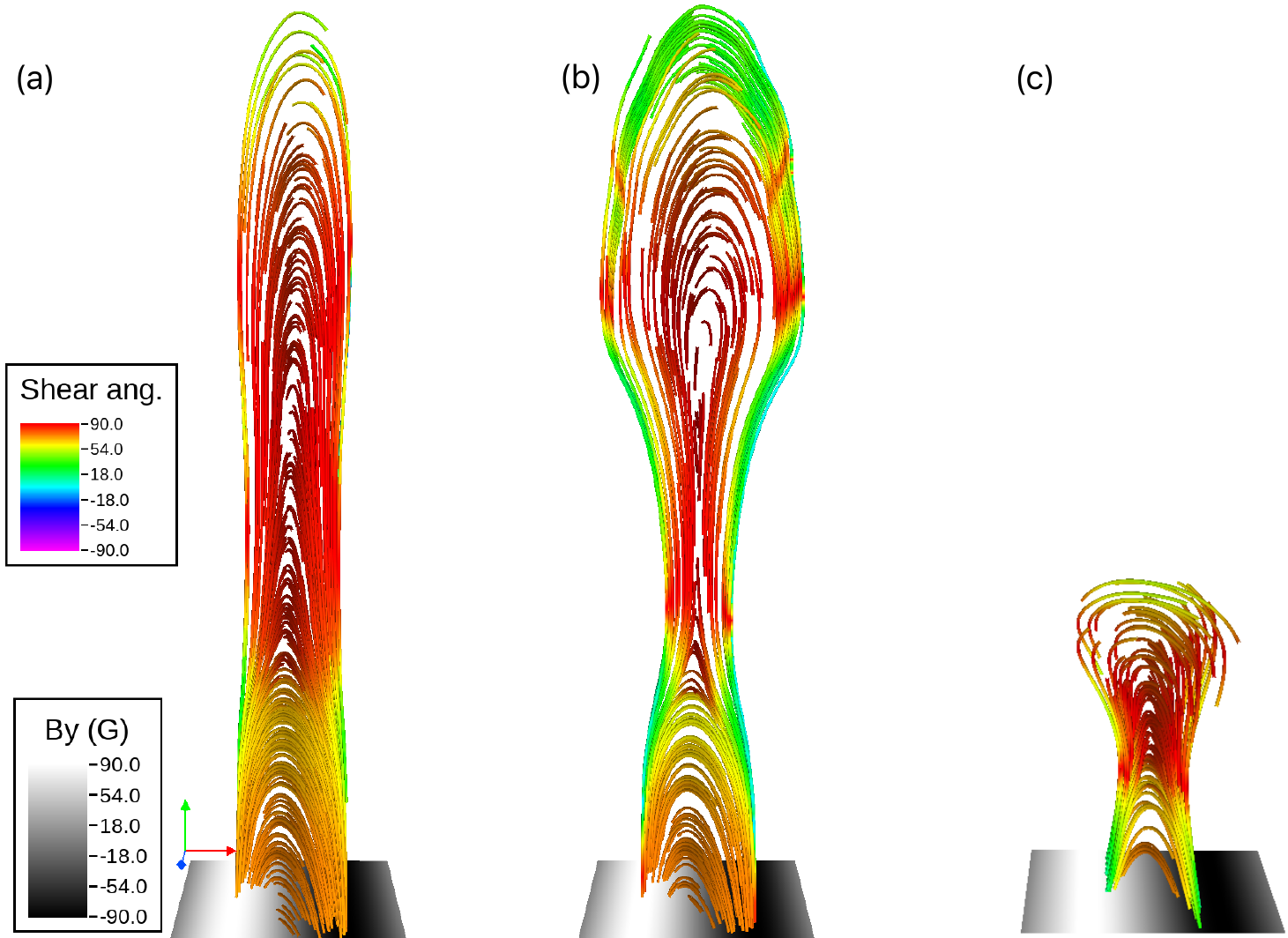}
\includegraphics[width=0.9\linewidth]{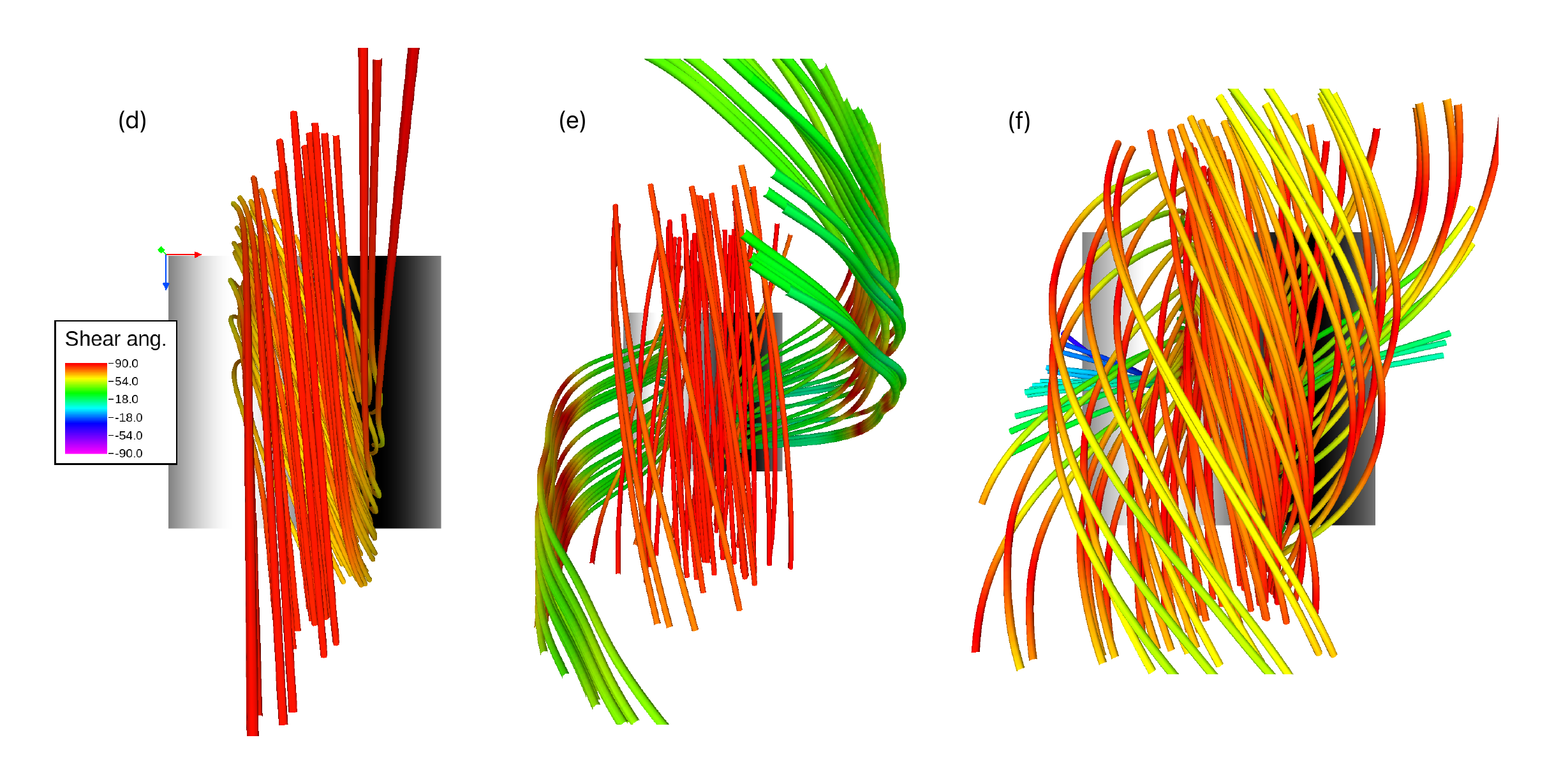}
\caption{Variation of the shear angle along the magnetic field lines from the central arcade region for $t=5.44$, 7.51 and 12.59 min going from left to right columns respectively, obtained from ``simulation 1''. The shear angle (in degree) are shown in the respective color bars. The top and bottom rows represent the side and top views of the field lines configuration respectively. The 2D slices show spatial variation of $B_y$ in the $x-z$ plane at the base of the simulation box, and values of $B_y$ is shown by the color bar common for all the panels. The orientation of the $x, y, z$ axes are shown in red, green and blue arrows respectively. The number of field lines at the bottom panels are chosen less compared to the corresponding top panels for a better visualization of the field lines.}
\label{fig:3DFL}
\end{figure*}

\begin{figure*}
    \centering
    \begin{subfigure}{0.8\columnwidth}
        \includegraphics[width=1\linewidth]{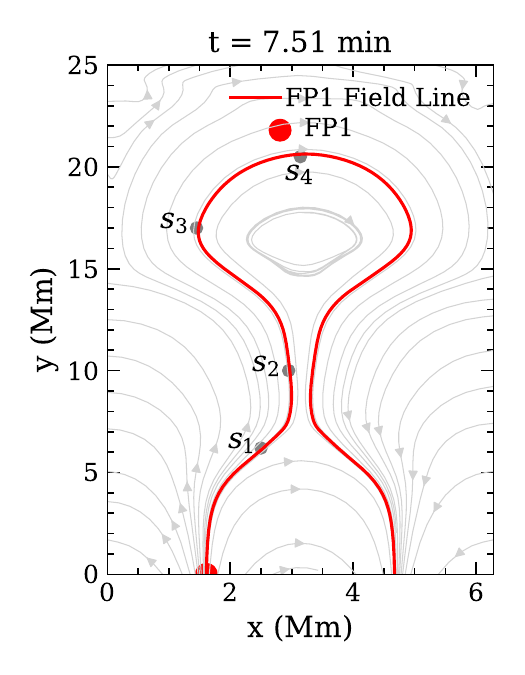}
        \caption{}
        \label{fig3_tl}
    \end{subfigure}
    \begin{subfigure}{0.8\columnwidth}
        \includegraphics[width=1\linewidth]{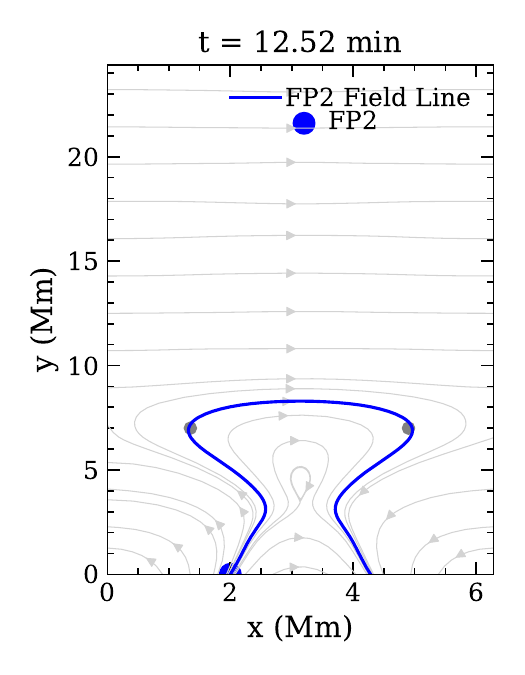}
        \caption{}
        \label{fig3_tr}
    \end{subfigure}
    
    \begin{subfigure}{0.8\columnwidth}
        \includegraphics[width=1\linewidth]{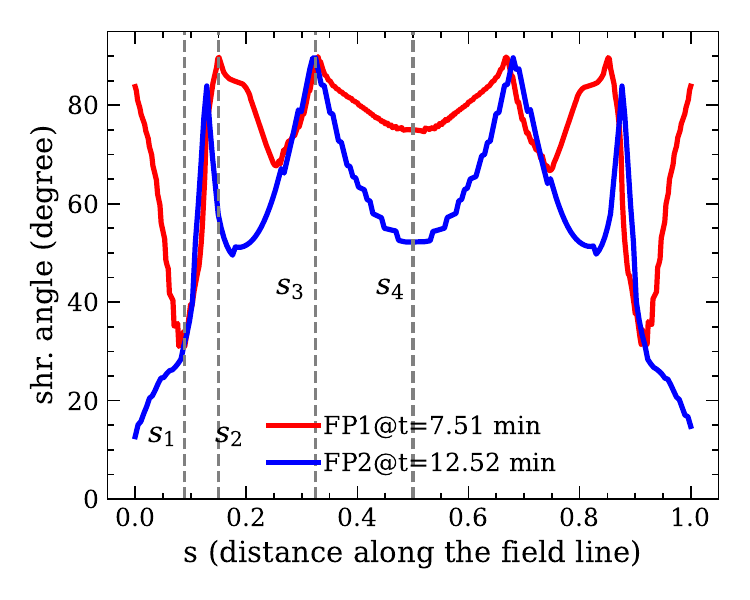}
        \caption{}
        \label{fig3_bl}
    \end{subfigure}
    \begin{subfigure}{0.8\columnwidth}
        \includegraphics[width=1\linewidth]{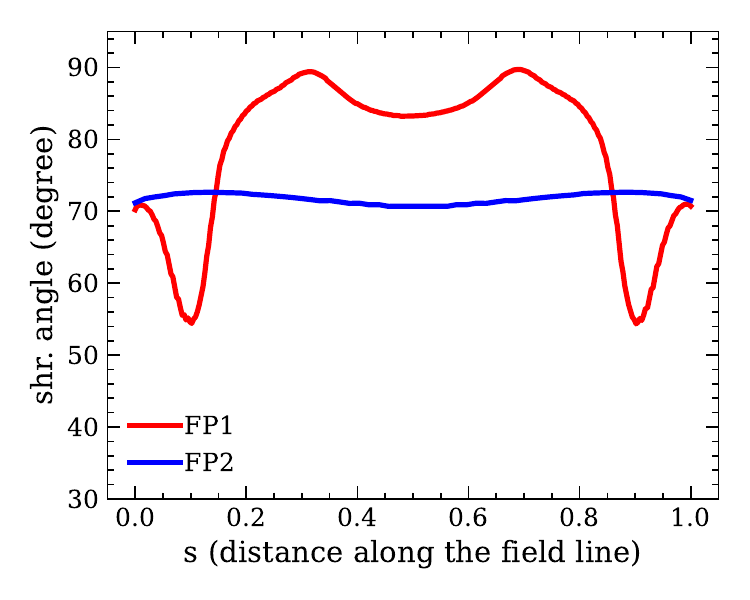}
        \caption{at $t=5.44$ min}
        \label{fig3_br}
    \end{subfigure}
    \caption{(a) Distribution of magnetic field lines (in gray) for the entire spatial domain of ``simulation 1'', at $t=7.51$ min, where the red curve represents the field line that is anchored at the seed point (foot point) location $(x, y)=(1.61, 0)$ Mm (FP1) as marked by the red circle. (b) Same as the top-left panel, but for $t=12.52$ min, where the blue field line is anchored at the seed point location $(x, y)=(2, 0)$ Mm (FP2). (c) The red and the blue curves represent the distribution of the shear, $\gamma$ along the red and blue field lines as shown in the top-left and top-right panels respectively, where the vertical dashed lines are the positions along the red field line marked by the gray dots ($s_1, .., s_4$) at the top-left panel. (d) Distribution of $\gamma$ along the field lines anchored at FP1 (red curve) and FP2 (blue curve) at the beginning of the semi-equilibrium phase at $t=5.44$ min.}
    \label{fig:shear_alongFL}
\end{figure*}

\begin{figure*}[hbt!]
    \centering
    \begin{subfigure}{1\columnwidth}
        \includegraphics[width=1\linewidth]{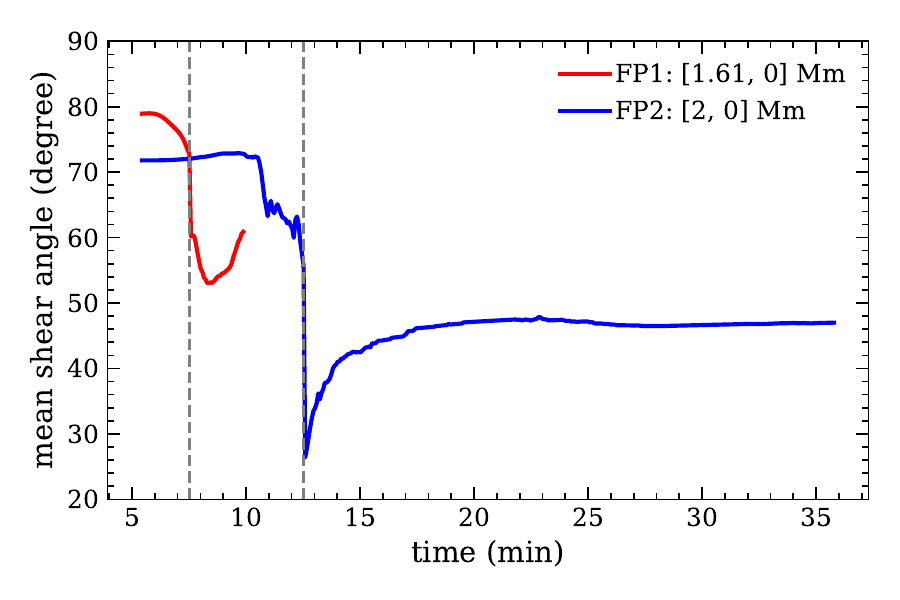}
        \caption{``simulation 1'': strong shear case}
        \label{fig4a}
    \end{subfigure}
    \begin{subfigure}{1\columnwidth}
        \includegraphics[width=1\linewidth]{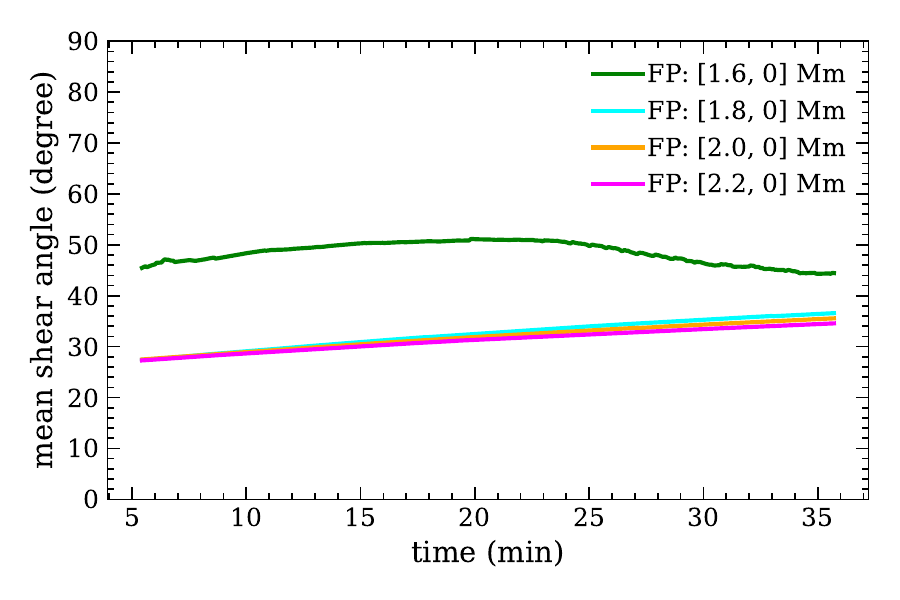}
        \caption{``simulation 2'': weak shear case}
        \label{fig4b}
    \end{subfigure}
    \caption{(a) Temporal variation of mean shear ($\bar{\gamma}$) of the field lines anchored at FP1 (red curve) and FP2 (blue curve) for the simulation with strong initial shear (``simulation 1''), which has multiple flux rope eruptions. The vertical dashed lines are the time markers at $t=7.51$ and 12.52 min, after which the reconnection occurs at those field lines. (b) Temporal variation of $\bar{\gamma}$ for the field lines anchored at different FPs as shown in the legend for the simulation with a weak initial shear (``simulation 2''), which do not have formation of any flux rope.}
    \label{fig:shear_evol}
\end{figure*} 

\begin{figure}
    \centering
   \includegraphics[width=0.95\linewidth]{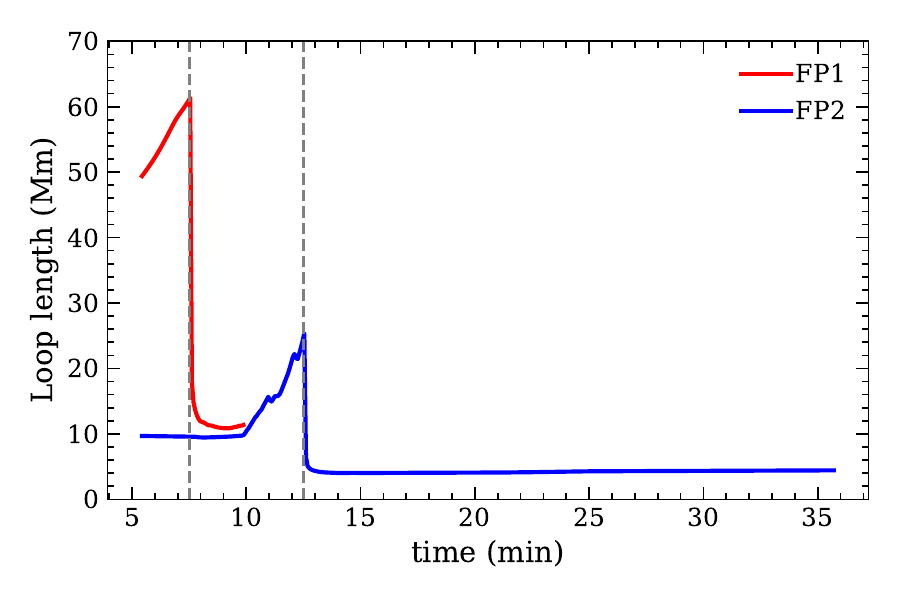}
   \caption{Temporal variation of the total arcade length anchored at FP1 (red curve) and FP2 (blue curve) for ``simulation 1''. The left and right vertical dashed lines are the time markers for $t=7.51$ and 12.52 min respectively.}
   \label{fig:fieldline_length}
\end{figure}  

\begin{figure}
   \centering
   \includegraphics[width=0.95\linewidth]{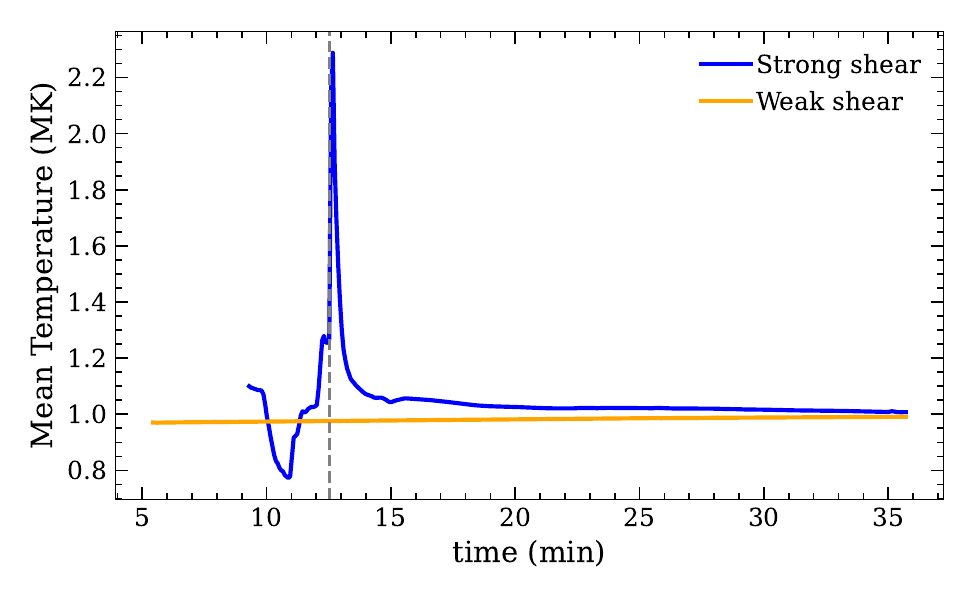}
   \caption{Time evolution of mean temperature, $\bar{T}$, along the field line that is anchored at FP2 ($x=2$ Mm, $y=0$). The blue and the orange curves denote the evolution for the strong (`simulation 1') and weak (`simulation 2') shear cases respectively. The vertical dashed line marks the time $t=12.52$ min.}
   \label{fig:fieldline_temp}
\end{figure}

\begin{figure*}[hbt!]
    \centering
   \includegraphics[width=0.49\linewidth]{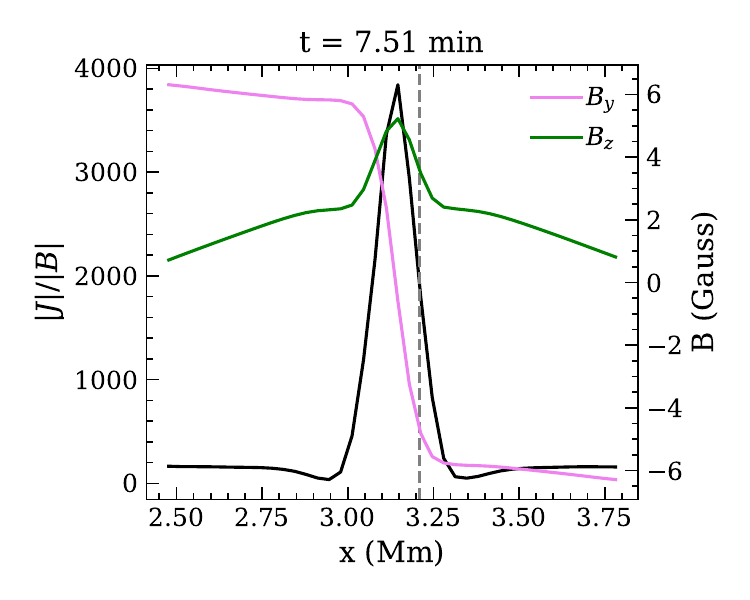}
   \includegraphics[width=0.49\linewidth]{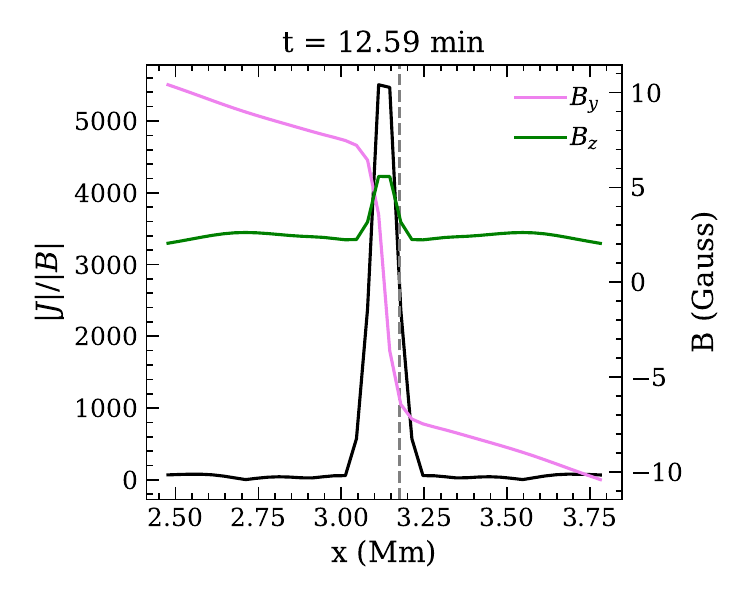}
    \caption{Spatial distribution of $\bf{|J|/|B|}$ (black), vertical component (magenta), and guide field component (green) along the horizontal direction that passes through the CS at $y=10$ Mm, and $y=3$ Mm for the first ($t=7.51$ min; left figure) and second ($t=12.59$ min; right figure) eruption phases respectively.}
    \label{fig:Rcs_loc}
\end{figure*} 

\begin{figure*}[hbt!]
    \centering
   \includegraphics[width=0.45\linewidth]{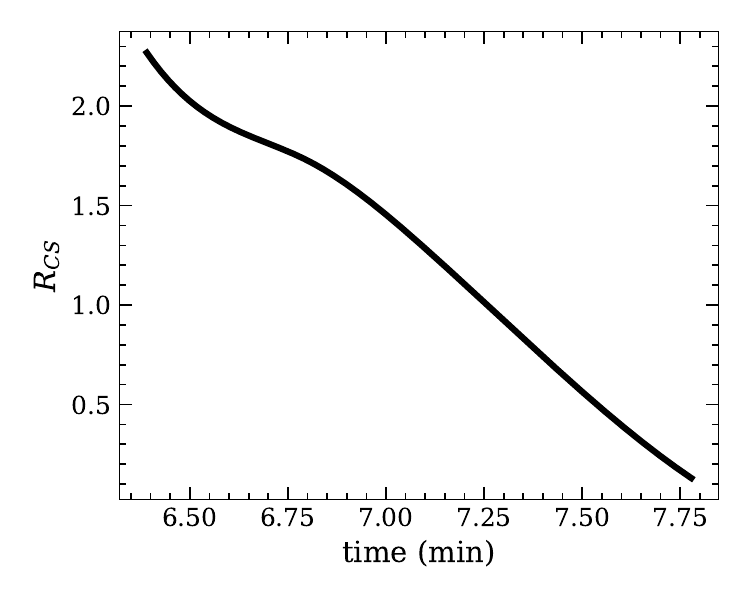}
   \includegraphics[width=0.45\linewidth]{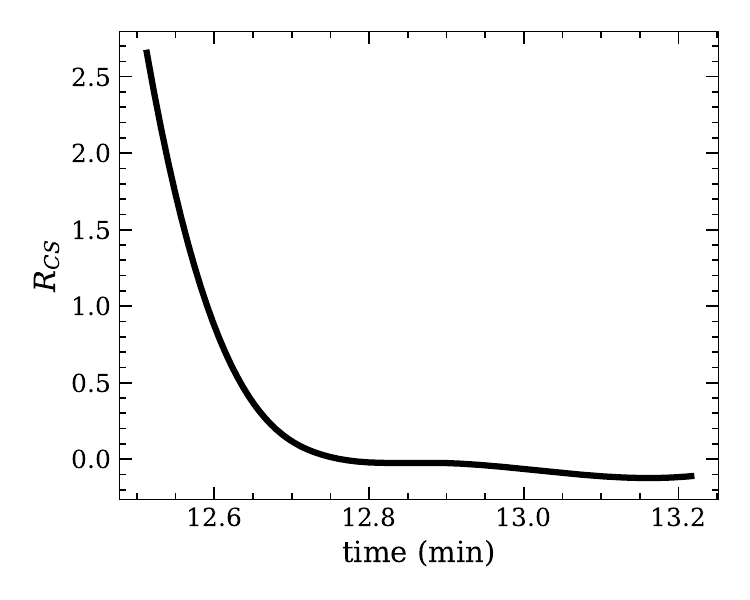}
    \caption{Temporal variation of the guide to vertical magnetic field component ratio, $R_{CS}\ (=-B_z/B_y)$ at the upstream location of CS marked by the vertical dashed lines in the left and right panels of Fig.~\ref{fig:Rcs_loc} during the first (left panel) and second (right panel) eruption phases respectively.}
    \label{fig:Rcs}
\end{figure*} 

For the ``simulation 1'', when the system reaches the semi-equilibrium state at $t=5.44$ min, the guide field component ($B_z$) confined within the loops contributes to the magnetic shear. This leads to the  stretching of the upper part ($y \gtrsim 13$ Mm) of the central arcade. Whereas, the lower part ($12\ \rm{Mm} \lesssim y \lesssim 7 \ \rm{Mm}$) squeezes along the polarity inversion line (PIL) that is present at $x=3.14$ Mm at the bottom boundary. The squeezing occurs due to the stretching of the straddling (half-) arcades that are present at the side boundaries (see Figs. \ref{fig1_tl}, and \ref{fig1_tm}). We thus abandon any attempt at describing the evolution of an ``exact'' isolated loop in particular, but, address the generic behavior of a coronal loop that is sandwiched and interacting between two side arcades as reported in different local box \citep[][and references therein]{1988:Mikic, 2016:sanjay, 2024:Lu}, and global \citep[][and references therein]{2006:Ding, 2009:soenen} models. 

We show the spatial distribution of the magnetic field lines projected in the $xy$ plane (top row), and the absolute value of the guide field component (bottom row) in Fig.~ \ref{fig:cs_FL_pmag}, for three different times: at the beginning of the semi-equilibrium phase ($t=5.44$ min), and the two eruption phases at $t=7.51$ and $12.59$ min respectively. The central part of the arcade leads to the formation of a vertical current sheet (CS) when the field lines ($B_y$ component with opposite polarity) come in close proximity to each other near the PIL as seen in other flux rope eruption models \citep[e.g.,][]{2020:wenzhi, Zhang:2021, Dahlin:2022}, and inferred in the flare and CME observations \citep{Reva:2016, Yan:2018, Patel:2020}. Concurrently, to trace the development of current sheets, we show the spatial distribution of $\bf{|J|/|B|}$ as a proxy (where, $\bf{J}$ is the electric current density), in the top row in Fig.~\ref{fig:cs_FL_pmag}. Here, eventual developments of the CSs around $x=3.14$ Mm can be seen by the dark gray regions in the Fig.~\ref{fig1_tm} and \ref{fig1_tr}. 

After the semi-equilibrium phase at around $t=7$ min, the field lines that are in close proximity to the vertical CS region allow a reconnection in a tether-cutting fashion \citep{Antiochos:1999, Moore:2001}. The presence of the guide field in those CS regions indicates a 2.5D reconnection (unlike a purely 2D reconnection), also reported in \cite{Sen-FMI:2025} regarding plasmoid instability in a coronal CS. This process adds twist to the field lines above the reconnection point to form a magnetic flux rope, and increases the magnetic tension of the underlying loops. This process separates the flux rope from the underlying arcades, and leads to the first eruption at around $t=7.51$ min, shown in Fig.~\ref{fig1_tm}. The flux rope holds the shear flux trapped within it (Fig.~\ref{fig1_bm}) and transported through the flux rope eruption. As a consequence of the reconnection, the post reconnection loops shrinks down carrying the shear flux within it, and contributes flux to the underlying arcades. 

Afterwards, due to the periodic condition at the horizontal boundaries, a pair of (straddling) flux ropes are also formed at $t\approx 10$ min at the side boundaries (shown in Figs. 1(c) and 1(d) in S24), those erupt and leave the simulation box at $t\approx 11.8$ min (see the associated animation with Fig. 1 in S24). During eruption of these (side) flux ropes, nearly horizontal magnetic fields are created at the upper and the lower parts of these erupting ropes, that have the opposite and same directions respectively to the underlying central arcade. The interaction of these side ropes with the remaining central arcade gradually leads to the formation of another flux rope, and leads to the second eruption (from the central arcade region) at around $t=12.59$ min (shown in Fig.~\ref{fig1_tr}) in a similar fashion like the first eruption. The interaction of the lateral arcades can be seen from the high $|\textbf{J}| / |\textbf{B}|$ regions in Figs.~\ref{fig1_tm}, and \ref{fig1_tr}. The study by \cite{cheng:2001} also demonstrated the similar fact that a continuous increase of magnetic shear in the underlying arcades can trigger reconnection process and leads to the formation of magnetic islands. After the second eruption phase, we do not see any more flux rope formation at the central arcade region, as there is no sufficient shear in the underlying loops that can lead to the process.

To appreciate the spatial orientation of the magnetic field lines, and distribution of shear along the magnetic field lines, we present the 3D views of the field lines, shown in Fig.~\ref{fig:3DFL}. The top and bottom rows display the side and top views of the configurations, respectively, at $t=5.44$, 7.51, and 12.59 min, progressing from left to right columns. In the Fig.~\ref{fig:3DFL}(a), the highly sheared segments of the central arcade, highlighted in red, become squeezed towards the PIL at $t=7.51$ min (Fig.~\ref{fig:3DFL}(b)). In the Fig.~\ref{fig:3DFL}(b), and (c), the portions of the field lines located near the PIL that are marked in red are highly sheared magnetic structures, and nearly parallel to the PIL. These magnetic structures contribute to the formation of channels which are capable of supporting filament materials as reported in \cite{Zhao:2015, Kinizhink:2015, Kinizhink:2017}, and references therein. These sheared portions of the field lines also play a role in the development of vertical current sheets (CSs), which form beneath the flux ropes during the eruption phases at $t=7.51$ and 12.59 min (see Figs. \ref{fig1_tm}, and \ref{fig1_tr}). The corresponding top views at $t=5.44$, 7.51, and 12.59 min are shown in the bottom row, where the concentration of highly sheared field lines near the PIL are evident.

Further to investigate the distribution of the shear angle along the field lines qualitatively, we trace magnetic field lines from selected foot points (FPs) at the bottom boundary and estimate the shear angle along those field lines. This is obtained by integrating the magnetic field line equation by a third-order Runge-Kutta method from a foot point until it reaches a boundary of the simulation domain, where the integration terminates. We also set a marker in the field line tracer routine to verify whether the integration terminates at the bottom boundary after tracing the field line from a given foot point (i.e., the seed point at the bottom boundary), indicating a closed loop. Conversely, if the field line tracer terminates at any of the side or top boundaries, it is identified as an open loop. We present the selected field lines shown in red and blue, for $t=7.51$ min and $12.52$ min in the Fig.~\ref{fig3_tl} and \ref{fig3_tr} respectively, where the FPs (or, seed points marked in red and blue circles) are located at the bottom boundary ($y=0$) at $x=1.61$ and 2 Mm, which we call `foot point 1' (FP1) and `foot point 2' (FP2) respectively, hereafter. The motivation of selecting these two foot points is to trace field lines from the central arcade region those reconnect during the first and second eruption phases respectively. We use a spatial co-ordinate, $s$, which is defined as the distance between the specific FPs (namely, FP1 and FP2) and an arbitrary position at the corresponding field lines. 

The distribution of the shear angle along the field lines anchored at FP1 and FP2 are shown in Fig.~\ref{fig3_bl} for two time instants of first and the second eruption phases. Here, we normalized the total length of the field lines to unity. We notice that the shear along the field lines is distributed in a quite inhomogeneous fashion, and with a (nearly) left-right symmetry from the apexes (i.e., at $s=0.5$) of the respective loops for both the eruption phases. To map the one-to-one correspondence of shear at different locations along the field line for the first eruption phase, we select four locations namely, $s_1, s_2, s_3,$ and $s_4$ as marked by the gray dots in the Fig.~\ref{fig3_tl}. 

We notice a strong shear ($\approx 80^\circ$) at the FP1 (at the location $s_0$) as marked by the red circle in Fig.~\ref{fig3_tl}. As we move along the field line from $s_0$, the shear falls sharply to $\approx 30^\circ$ at $s_1$ ($s=0.1$) as a consequence of predominance of the horizontal component over its guide field. The magnetic field is approximately in a relaxed state just before the flux rope formation, and follows a nearly magnetostatic equilibrium, so the $B_z(A_z)$ (where, $A_z$ is the $z$-component of the magnetic vector potential associated to $\bf{B}$). The constant $A_z$ also defines the field lines projected in the $x-y$ plane. This is also evident from the Figs. \ref{fig1_tm} and \ref{fig1_bm}, where the $B_z$ strength follows (nearly) the field lines. Therefore, the distribution of the magnetic shear along the FP1 field line is (mostly) governed by the $B_x$ component, as $B_z$ remains nearly uniform along the field line. Hence, smaller $B_x$ corresponds to higher shear, and higher $B_x$ corresponds to smaller shear along that field line.  

The shear reaches its maximum value ($\gamma=90^\circ$) at $s_2$ ($s \approx 0.15$), which is the upstream location of the CS, where the field lines are not yet reconnected, and mostly vertical when projected in the $x-y$ plane.  A similar behavior is noticed at $s_3$ ($s=0.32$), where the shear reaches its peak value, that corresponds to a location of an incipient flux rope, which is just about to form. Finally, at $s_4$, which is located at the apex of the incipient flux rope, shows a sufficiently strong shear of $\approx 75^\circ$. We repeat this calculation for the second eruption phase ($t=12.52$ min). In this case, we notice that the shears at the left and the right FPs ($x=2$, and 4.28 Mm respectively, at $y=0$) are weak ($\gamma \lesssim 15^\circ$), but comprise of stronger shear at the upper part of the loop, which reach to its maximum value of $\approx 90^\circ$ at $s\approx 0.32$ and 0.68 as marked by the gray dots in the Fig.~\ref{fig3_tr}, which are the parts of an incipient flux rope for the second eruption phase. 

The formation of the first flux rope occurs at a higher height than the second one (see Figs. \ref{fig1_tm} and \ref{fig1_tr}). To investigate the reason, we estimate the shear distribution at the beginning of the semi-equilibrium phase ($t=5.44$ min) along the field lines those are connected to the same two foot points, namely FP1 and FP2. For convenience, we call them FP1 field line, and FP2 field line hereafter, respectively. From Fig.~\ref{fig3_br}, we notice that the shear at the FP1 and FP2 are the same (around $70^\circ$), and the FP2 field line has a nearly uniform shear. On the other hand, the shear along the FP1 field line is much more inhomogeneous, and has a stronger shear at the upper part of the field line ($0.15 \lesssim s \lesssim 0.85$). Though the shear was spatially uniform at the initial stage ($t=0$), the occurrence of reconnections at the earlier time than the beginning of the semi-equilibrium stage ($t<5.44$ min, not shown) induces the inhomogeneity in shear distribution at the FP1 field line at the beginning of the semi-equilibrium phase. This higher shear at the upper part of the FP1 field line facilitates the formation of a flux rope at an earlier stage and at a higher height ($y \approx 17$ Mm) compared to the formation of flux rope from FP2 field line.     

Due to the strong inhomogeneity of shear that might appear along the field lines (as mentioned earlier), we estimate the mean shear along the field line ($\bar{\gamma}$), which we refer as mean shear for convenience, hereafter. This is estimated by integrating the shear angle along the field line and then dividing by the length of the field line,
\begin{align}
    \bar{\gamma} = \frac{\int \gamma \ \mathrm{d}s}{\int \mathrm{d}s}.
    \label{eq:mean_shear}
\end{align}

We calculate the temporal variation of $\bar{\gamma}$ for the FP1 and FP2 field lines, and present the result in the Fig.~\ref{fig4a}. At the semi-equilibrium phase, the mean shear of the FP1 field line is approximately $80^\circ$ and gradually decreases until $t=7.51$ min. During this period, the closed magnetic loop anchored to FP1 rises and stretches, corresponding to the elongation of the loop (shown by the red curve in Fig.~\ref{fig:fieldline_length}). Overall, the loop becomes more perpendicular to the PIL, which leads to the decrease of the non-potentiality, and hence the mean shear. The result is consistent with the study by \cite{2017:Qiu}, where they reported that the elongation of the loop, which was measured by the distance between the loop foot points, reduces the shear during a flare evolution. The mean shear angle reaches to $73^\circ$ at around $t=7.51$ minutes as indicated by the vertical dashed gray line in the Fig.~\ref{fig4a}. Shortly afterward, an upper part of the loop reconnects to form a flux rope at $t=7.58$ min, and the remaining bottom part shrinks downward forming an underlying arcade, which we call a post-reconnection loop. The shear flux carried by the flux rope is expelled upward during the eruption, resulting in a loss of free energy. The readers are referred to Fig. 1 in S24 for the time-series of the free to potential-field energy ratio capturing the time span of the first and second eruption phases, where we see that the ratio drops during those time periods. 

Due to the reconnection, the length of the closed field line (which is anchored at FP1) decreases from around 60 Mm at $t=7.51$ min to 10 Mm at $t\approx 8.3$ min (red curve in Fig.~\ref{fig:fieldline_length}). As a consequence, there is a sharp drop in the mean shear of the FP1 field line in the newly formed post-reconnection loop beneath the erupting rope to approximately $55^\circ$ at $t=8.3$ min. Meanwhile, the shear flux that is transported downward by the shrinking field lines contribute to the magnetic flux to the underlying loops. This enhances the non-potentiality of those underlying arcades, and consequently, the mean shear gradually increases between $t=8.3$  to 9.9 min  (Fig.~\ref{fig4a}). The calculation is not extended beyond this time, as the field line tracer traces a neighboring (open) field line from FP1 shortly afterward that belongs to a different loop family extending outward (to the left) of the central arcade.  

Similarly, we carry out the calculation for the field line that is anchored to FP2, and present the result by the blue curve in the Fig.~\ref{fig4a}. Here, we notice that the mean shear is around $72^\circ$ and remains nearly same from the start of the semi-equilibrium phase ($t=5.44$ min) until $t\approx 10$ min. In this phase the loop length remains nearly constant to $10$ Mm (blue curve in Fig.~\ref{fig:fieldline_length}). The tiny variations of the mean shear and loop length between $t\approx 11$ to 12 min is due to the flickering nature of the loop, which is caused by the periodic boundary condition and the straddling half loops at the sides. Shortly after $t=10$ min, during the onset of the second eruption phase, the loop length starts to increase reaching to around 25 Mm at $t\approx 12.52$ min, and lowering of the non-potentiality of the loop similar to the FP1 field line as described above. Therefore, the mean shear drops to around $60^\circ$ at that time. Right after $t=12.52$ min, when a part of that field line reconnects to form an erupting flux rope, we see a sharp drop of mean shear to $\approx 25^\circ$, and the field line length decreases to around 4 Mm. This length remains almost constant until the end of the evolution. 

The shear flux carried downward contributes to the flux in the underlying loops similar to the first post-eruption phase, and we see a gradual increment of the mean shear (blue curve at the Fig.~\ref{fig4a}). This value reaches to $\approx 48^\circ$ at $t \approx 20$ min. After this stage, there is no more significant shear flux injected to the underlying loop, and therefore the mean shear remains almost constant until the end of the evolution. The amount of shear remains at the central loops after this, is not sufficient to form flux ropes from the central arcade region. 

On the other hand, we investigate the temporal variation of the mean shear for ``simulation 2'' which has a weaker (spatially uniform) shear ($\gamma = 25.8^\circ$) than ``simulation 1'' ($\gamma = 72.5^\circ$) at the initial time. To do that, we trace the field lines of the central loop from different foot point locations of $(x, y)=[(1.6, 0), (1.8,0), (2,0), (2.2,0)]$ Mm, and estimate $\bar{\gamma}$ similarly as described above. We notice that the mean shear angle of the field line from FP location (1.6, 0) Mm is maximum which is around $45^\circ$, and remains nearly constant between the same time span of the evolution of ``simulation 1'' as shown in the Fig.~\ref{fig4b}. However, we notice that field lines those are anchored with the other FPs are closer to the PIL (at $x=3.14$ Mm) have weaker shear, and $\bar{\gamma}$ varies between $\approx 28^\circ$ to $36^\circ$ during the evolution. In this case, the shear present in any of the field lines of the central arcade region is not sufficient to trigger the formation of flux rope, and remains in a quasi-stationary state till the end of the evolution. 

To investigate the temperature evolution of the pre- and post-reconnection loops, we estimate the mean temperature along the field line similar to the $\bar{\gamma}$ in Eq. \ref{eq:mean_shear}, i.e.,
\begin{align} \label{eq:T_mean}
    \bar{T} = \frac{\int T \rm{d}s}{\int \rm{d}s}.
\end{align}
The result is presented in Fig.~\ref{fig:fieldline_temp}. Here, we focus on the results pertaining specifically to the second eruption phase. To this end, we select magnetic field lines anchored at FP2 (i.e., $[x ,y]=[2,0]$ Mm) in order to estimate the evolution of the mean temperature, $\bar{T}$, capturing both the pre- and post-eruption temporal domains. In the case of weak shear (`simulation 2'), no significant change is observed in the temperature ($\bar{T}$) evolution, which remains at around 1 MK (which is same as the initial iso-thermal condition) throughout the evolution. In contrast, for the strong shear case (`simulation 1'), $\bar{T}$ exhibits substantial variation, ranging approximately from 0.8 to 2.3 MK. 

During the interval between $t=9.3$ to 11 min, the temperature decreases from about 1.1 MK to 0.8 MK. During this pre-reconnection phase, the stretching of the loop leads to an increase of thermal conduction timescale (which varies $\sim L^2$, where, $L$ is the characteristic length scale of the loop). Which means the thermal conduction becomes less efficient than a shorter length scale scenario. Therefore, the thermal conduction can not react faster to enhance the temperature than the cooling mechanism that facilitates mostly due to adiabatic expansion of the loops. Followed by that, $\bar{T}$ rises to approximately 2.3 MK at $t \approx 12.5$ min, primarily due to ohmic heating in the region of enhanced current density at the vertical CS (see Fig.~\ref{fig1_tr}). Shortly thereafter, at $t=12.6$ min, the mean temperature of the post-reconnection loop begins to decrease again, reaching to the initial temperature of around 1 MK by $t\approx15$ min, and remaining at that level through the end of the evolution at around $t=35$ min. This temperature decline is attributed to the downward transport of thermal energy along the loop due to thermal conduction. Notably, the fast temperature change (with timescale $\sim$ few minutes) during the transition phase from the reconnection to post-reconnection period (between $t=12.6$ to 15 min) is governed by the small length scale of the loop ($\sim$ few Mm), which corresponds to small thermal conduction timescale, as well as the temperature inhomogeneities present along the loop.        

The relative strength between the reconnection field and the guide field components at the upstream location in a CS can be a deciding factor about the particle acceleration and bulk heating \citep{Dahlin:2022, 2023ApJ...955...34Q}. To assess this, we estimate the time evolution of the ratio between the guide and vertical magnetic field components, defined as $R_{CS} = -B_z / B_y$, at the right side ($x > 3.14$ Mm) of the upstream region of the CS during both eruption phases (at $t = 7.51$ and 12.59 min). At this upstream location, $B_y$ consistently remains negative in our case (i.e., directed downward), whereas $B_z$ can vary between positive and negative values. Thus, the adopted convention of $R_{CS}$ helps to identify when the guide field component at this location changes the polarity. To perform this analysis, we first select a vertical span encompassing the CS that is present at $x = 3.14$ Mm and identify the pixel within this region that corresponds to the maximum value of $\bf{|J| / |B|}$. This is also crosschecked with the maps in Figs. \ref{fig1_tm} and \ref{fig1_tr}. We then take a horizontal cut of the quantity, $\bf{|J| / |B|}$ that passes through to that specific pixel with peak $\bf{|J| / |B|}$, as shown by the black curves in Fig.~\ref{fig:Rcs_loc} for $t = 7.51$ and 12.59 min, respectively. To determine the upstream locations on the right side of the evolving CS, we select the pixels that reach 50\% of the instantaneous peak value of $\bf{|J| / |B|}$ while moving right along the $x$-direction from $x = 3.14$ Mm. These upstream positions are marked by vertical gray dashed lines in the left and right panels of Fig.~\ref{fig:Rcs_loc} for the first and second eruption phases, respectively. The reconnecting ($B_y$) and the guide ($B_z$) field profiles are also over plotted to show the consistency of the CS structure. Using this automated technique, we estimate the temporal evolution of $R_{CS}$ and present the results in Fig.~\ref{fig:Rcs}. 

Here, the left and right panels correspond the time span that captures the first and second eruption phases, respectively. During both the pre-eruption phases, $R_{CS}$ is more than unity, indicating that the guide field is stronger than the reconnection field component. In these periods, plasmoid-driven particle acceleration should be suppressed, and corresponds to a less production of nonthermal electrons due to a weakened Fermi-type mechanism as suggested in \citep{Arnold:2021}. However, during the late phase of eruption, when $R_{CS}$ decreases towards zero, the reconnection field component becomes dominant over the guide field. At this stage, the energy contribution from non-thermal electrons exceeds that of hot thermal electrons, resulting in efficient particle acceleration as proposed by \citep{Dahlin:2016}. This suggests that during the pre-eruption phases, the dominant guide field primarily drives bulk heating, whereas during the late phases of eruptions, the dominant reconnection field component facilitates the particle acceleration more efficiently. It also gives a potential direction to explain the ``hot onset'' phenomena, where the bulk heating is observed prior to the efficient particle acceleration in flares as reported in \cite{Hudson:2021}.

\section{Summary and discussion}\label{sec:summary}
We summarize our key findings and their significance in the following.

We present results from a 2.5D MHD model of multiple (homologous) flux rope eruptions (S24), focusing on the role of magnetic shear distribution in facilitating these eruptions. Specifically, we analyze two cases by varying the initial magnetic shear ($\gamma$), while keeping all other parameters constant. In both cases, the system initially (at $t=0$) exhibits mechanical imbalance. However, it transiently settled into a nearly force-free and thermally balanced semi-equilibrium state after approximately $5.44$ min. Shortly after that, for the strong shear case ($\gamma=72.6^\circ$: ``simulation 1''), the system produces two flux ropes from the central arcade region those erupt at around 7.51 and 12.59 min. In contrast, the weak shear case ($\gamma=25.8^\circ$: ``simulation 2'') does not show any flux rope formation after it reaches the semi-equilibrium phase, and remains in a quasi-stationary state thereafter. These results suggest that magnetic shear in the arcades due to presence of the guide field, plays an intrinsic role in formation and eruption of flux ropes. In this work, we focus on the formation and eruption of the flux ropes only from the central arcade region, and do not pay attention to the (partial) flux ropes those are formed at the side boundaries due to the initial magnetic field prescription and periodic boundary condition (see S24 for more details).  

The initial shear was spatially uniform in the entire simulation domain, but a nonuniformity is developed in the field lines at the beginning of the semi-equilibrium phase, as the stage before the semi-equilibrium phase ($t < 5.44$ min) consists of reconnections (not shown) due to the presence of Lorentz force. Formation of the first flux rope occurs at a higher height than the second one, because, at the beginning of the semi-equilibrium phase, the field lines that are connected with the outer legs (away from the PIL) of the arcade consist of more shear at a higher height than the field lines from the inner legs. The field lines in the vicinity of the vertical CS are highly sheared, and nearly parallel to the PIL. These are the parts of the field lines those reconnect. This implies that the amount of shear flux that is ejected upward by the eruptive flux ropes, and the residual flux that is transported downward to the post reconnection loops is decided by the location of reconnection on the arcade.  

During the pre-eruption phase, a gradual decrease in the mean shear ($\bar{\gamma}$) of the field lines is associated with the stretching of magnetic loops, which continues until magnetic reconnection occurs. Following the reconnection event, part of the arcade structure forms a flux rope that erupts and carry the shear flux with it, while the remaining portion of the arcade becomes the post-reconnection loop. Consequently, the mean shear of the reconnecting loop drops drastically just after the reconnection. In other words, a sudden drop of mean shear of a flaring loop may indicate a reconnection event. After the eruption, the downward-transported residual shear flux contributes to the magnetic flux in the underlying arcades, leading to a gradual increment in non-potentiality. Observationally, estimating mean shear along field lines is extremely challenging, if not impossible. However, temporal variations of magnetic shear at the photosphere are measurable and have been reported for various flare events. For example, according to the study by \cite{Gosain:2010}, the horizontal shear (referred to as ``dip shear'') at the sunspot penumbra regions decreases following an X-class flare. \cite{Petrie:2019} reported a reduction in shear during a flare in an active region. Moreover, \cite{Magara:2009} found that shear decreases even before the onset of an X-class flare. Though the trend of the mean shear obtained from our study, and the photospheric shear for different flare observations follows similar pattern, the one-to-one correspondence between these two quantities cannot be established across different phases of the eruption. Still the similar trends between our study and the existing observations encourage further investigations to explore any potential correlation between these two.

This work highlights the possibility of fast temperature evolution (timescale $\sim$ few minutes) during the pre-reconnection, and the transition period between the reconnection to post-reconnection phases for small-scale flare-associated loops. Our findings thus motivate the observational investigations aimed at confirming this scenario.

This study also sheds light in the understanding of energy release processes during an eruptive flare. According to the studies by \cite{Dahlin:2016, Arnold:2021}, it is demonstrated that the strong guide field suppresses the plasmoid-driven particle acceleration. The bulk heating on the other hand, which is the increase in thermal energy of the plasma (raising the temperature of the entire ion/electron population) is (mostly) insensitive to the guide field. Therefore, we expect that the plasma heating should dominate over the particle acceleration during the onset phase of the eruptions when the strength of guide field dominates over the reconnection field component, which might serve as a potential clue for a better understanding of the "hot onset" phenomena in flare observation \citep{Hudson:2021}. On the other hand, during the late phase of the eruption stage, when the reconnection field component dominates over the guide field, we expect a more efficient particle acceleration.

This work primarily highlights the role of magnetic shear distribution in the formation and eruption of multiple flux ropes within a 2.5D Cartesian geometry. The eruptions in our model emerges without any driving agent beyond the initial (mechanically imbalanced) conditions, unlike e.g., \cite{Dahlin:2022}, where shear flow was injected at the bottom boundary in a relaxed system. However, an extension of this work in a 3D geometry starting from the equilibrium configuration for a local box simulation with coupling to the lower atmospheres, and for a large-scale global model can be an interesting aspect to investigate in future. Nevertheless, the current study demonstrates that the distribution and variability of guide field is an inevitable quantity to trigger the formation and eruption of flux ropes, and energy release (through bulk heating and particle acceleration) in the Sun, which are essential to resolve the broader aspect coronal heating mystery.

\begin{acknowledgements}
SS acknowledges support by the European Research Council through the Synergy Grant \#810218 (``The Whole Sun”, ERC-2018-SyG). He thankfully acknowledges the technical expertise and assistance provided by the Spanish Supercomputing Network (Red Espa\~{n}ola de Supercomputaci{\'o}n), as well as the computer resources used: the LaPalma Supercomputer, located at the Instituto de Astrof{\'i}sica de Canarias. We would like to thank the anonymous referee for the insightful and useful comments that improved the manuscript considerably. We would like to thank Joel Dahlin, David Pontin, and James Klimchuk for useful discussions. Data visualization and analysis were performed using \href{https://www.python.org/downloads/release/python-3119/}{Python 3.11.9}, \href{https://yt-project.org/}{yt-project}, \href{https://visit-dav.github.io/visit-website/index.html}{Visit 2.10}, and \href{https://sgpearsevapor.readthedocs.io/en/latest/}{Vapor 3}. SSN acknowledges the supports from NSF-AGS-1954503, NASA-LWS-
80NSSC21K0003, and 80NSSC21K1671 grants and the I+D+i project PID2023-147708NB-I00 funded by MICIU/AEI/10.13039/501100011033/ and by FEDER, EU. 
\end{acknowledgements}  

\bibliographystyle{aa} 
\bibliography{reference}  

\begin{thebibliography}{61}
\expandafter\ifx\csname natexlab\endcsname\relax\def\natexlab#1{#1}\fi

\bibitem[{{Antiochos}(1998)}]{1998ApJ...502L.181A}
{Antiochos}, S.~K. 1998, \apjl, 502, L181

\bibitem[{{Antiochos} {et~al.}(1999){Antiochos}, {DeVore}, \& {Klimchuk}}]{Antiochos:1999}
{Antiochos}, S.~K., {DeVore}, C.~R., \& {Klimchuk}, J.~A. 1999, \apj, 510, 485

\bibitem[{{Arnold} {et~al.}(2021){Arnold}, {Drake}, {Swisdak}, {Guo}, {Dahlin}, {Chen}, {Fleishman}, {Glesener}, {Kontar}, {Phan}, \& {Shen}}]{Arnold:2021}
{Arnold}, H., {Drake}, J.~F., {Swisdak}, M., {et~al.} 2021, \prl, 126, 135101

\bibitem[{{Aulanier} {et~al.}(2012){Aulanier}, {Janvier}, \& {Schmieder}}]{Aulanier:2012}
{Aulanier}, G., {Janvier}, M., \& {Schmieder}, B. 2012, \aap, 543, A110

\bibitem[{{Cheng} \& {Choe}(2001)}]{cheng:2001}
{Cheng}, C.~Z. \& {Choe}, G.~S. 2001, Earth, Planets and Space, 53, 597

\bibitem[{{Cheng} {et~al.}(2003){Cheng}, {Ren}, {Choe}, \& {Moon}}]{cheng:2003}
{Cheng}, C.~Z., {Ren}, Y., {Choe}, G.~S., \& {Moon}, Y.~J. 2003, \apj, 596, 1341

\bibitem[{{Chifor} {et~al.}(2008){Chifor}, {Young}, {Isobe}, {Mason}, {Tripathi}, {Hara}, \& {Yokoyama}}]{2008A&A...481L..57C}
{Chifor}, C., {Young}, P.~R., {Isobe}, H., {et~al.} 2008, \aap, 481, L57

\bibitem[{{Dahlin} {et~al.}(2022){Dahlin}, {Antiochos}, {Qiu}, \& {DeVore}}]{Dahlin:2022}
{Dahlin}, J.~T., {Antiochos}, S.~K., {Qiu}, J., \& {DeVore}, C.~R. 2022, \apj, 932, 94

\bibitem[{{Dahlin} {et~al.}(2016){Dahlin}, {Drake}, \& {Swisdak}}]{Dahlin:2016}
{Dahlin}, J.~T., {Drake}, J.~F., \& {Swisdak}, M. 2016, Physics of Plasmas, 23, 120704

\bibitem[{{D{\'e}moulin} \& {Pariat}(2009)}]{Demoulin:2009}
{D{\'e}moulin}, P. \& {Pariat}, E. 2009, Advances in Space Research, 43, 1013

\bibitem[{{Ding} {et~al.}(2006){Ding}, {Hu}, \& {Wang}}]{2006:Ding}
{Ding}, J.~Y., {Hu}, Y.~Q., \& {Wang}, J.~X. 2006, \solphys, 235, 223

\bibitem[{{Gosain} \& {Venkatakrishnan}(2010)}]{Gosain:2010}
{Gosain}, S. \& {Venkatakrishnan}, P. 2010, \apjl, 720, L137

\bibitem[{{Hagyard} \& {Rabin}(1986)}]{1986AdSpR...6f...7H}
{Hagyard}, M.~J. \& {Rabin}, D.~M. 1986, Advances in Space Research, 6, 7

\bibitem[{{Hagyard} {et~al.}(1984){Hagyard}, {Smith}, {Teuber}, \& {West}}]{1984SoPh...91..115H}
{Hagyard}, M.~J., {Smith}, Jr., J.~B., {Teuber}, D., \& {West}, E.~A. 1984, \solphys, 91, 115

\bibitem[{{Hong} {et~al.}(2011){Hong}, {Jiang}, {Zheng}, {Yang}, {Bi}, \& {Yang}}]{2011ApJ...738L..20H}
{Hong}, J., {Jiang}, Y., {Zheng}, R., {et~al.} 2011, \apjl, 738, L20

\bibitem[{{Hudson} {et~al.}(2021){Hudson}, {Sim{\~o}es}, {Fletcher}, {Hayes}, \& {Hannah}}]{Hudson:2021}
{Hudson}, H.~S., {Sim{\~o}es}, P. J.~A., {Fletcher}, L., {Hayes}, L.~A., \& {Hannah}, I.~G. 2021, \mnras, 501, 1273

\bibitem[{{Inoue} {et~al.}(2011){Inoue}, {Kusano}, {Magara}, {Shiota}, \& {Yamamoto}}]{Inoue:2011}
{Inoue}, S., {Kusano}, K., {Magara}, T., {Shiota}, D., \& {Yamamoto}, T.~T. 2011, \apj, 738, 161

\bibitem[{{Keppens} {et~al.}(2012){Keppens}, {Meliani}, {van Marle}, {Delmont}, {Vlasis}, \& {van der Holst}}]{2012JCoPh.231..718K}
{Keppens}, R., {Meliani}, Z., {van Marle}, A.~J., {et~al.} 2012, Journal of Computational Physics, 231, 718

\bibitem[{{Keppens} {et~al.}(2003){Keppens}, {Nool}, {T{\'o}th}, \& {Goedbloed}}]{keppens2003}
{Keppens}, R., {Nool}, M., {T{\'o}th}, G., \& {Goedbloed}, J.~P. 2003, Computer Physics Communications, 153, 317

\bibitem[{{Keppens} {et~al.}(2023){Keppens}, {Popescu Braileanu, B.}, {Zhou, Y.}, {Ruan, W.}, {Xia, C.}, {Guo, Y.}, {Claes, N.}, \& {Bacchini, F.}}]{keppens2023}
{Keppens}, R., {Popescu Braileanu, B.}, {Zhou, Y.}, {et~al.} 2023, A\&A, 673, A66

\bibitem[{Keppens {et~al.}(2021)Keppens, Teunissen, Xia, \& Porth}]{keppens2021}
Keppens, R., Teunissen, J., Xia, C., \& Porth, O. 2021, Computers \& Mathematics with Applications, 81, 316, development and Application of Open-source Software for Problems with Numerical PDEs

\bibitem[{{Kliem} \& {T{\"o}r{\"o}k}(2006)}]{2006PhRvL..96y5002K}
{Kliem}, B. \& {T{\"o}r{\"o}k}, T. 2006, \prl, 96, 255002

\bibitem[{{Knizhnik} {et~al.}(2015){Knizhnik}, {Antiochos}, \& {DeVore}}]{Kinizhink:2015}
{Knizhnik}, K.~J., {Antiochos}, S.~K., \& {DeVore}, C.~R. 2015, \apj, 809, 137

\bibitem[{{Knizhnik} {et~al.}(2017){Knizhnik}, {Antiochos}, {DeVore}, \& {Wyper}}]{Kinizhink:2017}
{Knizhnik}, K.~J., {Antiochos}, S.~K., {DeVore}, C.~R., \& {Wyper}, P.~F. 2017, \apjl, 851, L17

\bibitem[{{Kumar} {et~al.}(2016){Kumar}, {Bhattacharyya}, {Joshi}, \& {Smolarkiewicz}}]{2016:sanjay}
{Kumar}, S., {Bhattacharyya}, R., {Joshi}, B., \& {Smolarkiewicz}, P.~K. 2016, \apj, 830, 80

\bibitem[{{Lu} {et~al.}(1993){Lu}, {Wang}, \& {Wang}}]{1993SoPh..148..119L}
{Lu}, Y., {Wang}, J., \& {Wang}, H. 1993, \solphys, 148, 119

\bibitem[{{Lu} {et~al.}(2024){Lu}, {Chen}, {Guo}, {Ding}, {Wang}, {Yu}, {Ni}, \& {Xia}}]{2024:Lu}
{Lu}, Z., {Chen}, F., {Guo}, J.~H., {et~al.} 2024, \apjl, 973, L1

\bibitem[{{MacTaggart} {et~al.}(2021){MacTaggart}, {Prior}, {Raphaldini}, {Romano}, \& {Guglielmino}}]{2021NatCo..12.6621M}
{MacTaggart}, D., {Prior}, C., {Raphaldini}, B., {Romano}, P., \& {Guglielmino}, S.~L. 2021, Nature Communications, 12, 6621

\bibitem[{{Magara}(2009)}]{Magara:2009}
{Magara}, T. 2009, \apj, 702, 386

\bibitem[{{McGlasson} {et~al.}(2019){McGlasson}, {Panesar}, {Sterling}, \& {Moore}}]{2019ApJ...882...16M}
{McGlasson}, R.~A., {Panesar}, N.~K., {Sterling}, A.~C., \& {Moore}, R.~L. 2019, \apj, 882, 16

\bibitem[{{Mikic} {et~al.}(1988){Mikic}, {Barnes}, \& {Schnack}}]{1988:Mikic}
{Mikic}, Z., {Barnes}, D.~C., \& {Schnack}, D.~D. 1988, \apj, 328, 830

\bibitem[{{Moore} {et~al.}(2001){Moore}, {Sterling}, {Hudson}, \& {Lemen}}]{Moore:2001}
{Moore}, R.~L., {Sterling}, A.~C., {Hudson}, H.~S., \& {Lemen}, J.~R. 2001, \apj, 552, 833

\bibitem[{{Moreno-Insertis} {et~al.}(2008){Moreno-Insertis}, {Galsgaard}, \& {Ugarte-Urra}}]{2008ApJ...673L.211M}
{Moreno-Insertis}, F., {Galsgaard}, K., \& {Ugarte-Urra}, I. 2008, \apjl, 673, L211

\bibitem[{{Nayak} {et~al.}(2024){Nayak}, {Sen}, {Shrivastav}, {Bhattacharyya}, \& {Athiray}}]{2024:Nayak}
{Nayak}, S.~S., {Sen}, S., {Shrivastav}, A.~K., {Bhattacharyya}, R., \& {Athiray}, P.~S. 2024, \apj, 975, 143

\bibitem[{{Panesar} {et~al.}(2016){Panesar}, {Sterling}, \& {Moore}}]{2016ApJ...822L..23P}
{Panesar}, N.~K., {Sterling}, A.~C., \& {Moore}, R.~L. 2016, \apjl, 822, L23

\bibitem[{{Panesar} {et~al.}(2018){Panesar}, {Sterling}, \& {Moore}}]{2018ApJ...853..189P}
{Panesar}, N.~K., {Sterling}, A.~C., \& {Moore}, R.~L. 2018, \apj, 853, 189

\bibitem[{{Patel} {et~al.}(2020){Patel}, {Pant}, {Chandrashekhar}, \& {Banerjee}}]{Patel:2020}
{Patel}, R., {Pant}, V., {Chandrashekhar}, K., \& {Banerjee}, D. 2020, \aap, 644, A158

\bibitem[{{Patty} \& {Hagyard}(1986)}]{1986SoPh..103..111P}
{Patty}, S.~R. \& {Hagyard}, M.~J. 1986, \solphys, 103, 111

\bibitem[{{Peter} {et~al.}(2013){Peter}, {Bingert}, {Klimchuk}, {de Forest}, {Cirtain}, {Golub}, {Winebarger}, {Kobayashi}, \& {Korreck}}]{2013:Peter}
{Peter}, H., {Bingert}, S., {Klimchuk}, J.~A., {et~al.} 2013, \aap, 556, A104

\bibitem[{{Petrie}(2019)}]{Petrie:2019}
{Petrie}, G. J.~D. 2019, \apjs, 240, 11

\bibitem[{{Porth} {et~al.}(2014){Porth}, {Xia}, {Hendrix}, {Moschou}, \& {Keppens}}]{2014ApJS..214....4P}
{Porth}, O., {Xia}, C., {Hendrix}, T., {Moschou}, S.~P., \& {Keppens}, R. 2014, \apjs, 214, 4

\bibitem[{{Qiu} {et~al.}(2023){Qiu}, {Alaoui}, {Antiochos}, {Dahlin}, {Swisdak}, {Drake}, {Robison}, {DeVore}, \& {Uritsky}}]{2023ApJ...955...34Q}
{Qiu}, J., {Alaoui}, M., {Antiochos}, S.~K., {et~al.} 2023, \apj, 955, 34

\bibitem[{{Qiu} {et~al.}(2017){Qiu}, {Longcope}, {Cassak}, \& {Priest}}]{2017:Qiu}
{Qiu}, J., {Longcope}, D.~W., {Cassak}, P.~A., \& {Priest}, E.~R. 2017, \apj, 838, 17

\bibitem[{{Reva} {et~al.}(2016){Reva}, {Ulyanov}, \& {Kuzin}}]{Reva:2016}
{Reva}, A.~A., {Ulyanov}, A.~S., \& {Kuzin}, S.~V. 2016, \apj, 832, 16

\bibitem[{{Ruan} {et~al.}(2020){Ruan}, {Xia}, \& {Keppens}}]{2020:wenzhi}
{Ruan}, W., {Xia}, C., \& {Keppens}, R. 2020, \apj, 896, 97

\bibitem[{{Schmieder} {et~al.}(1996){Schmieder}, {Demoulin}, {Aulanier}, \& {Golub}}]{1996ApJ...467..881S}
{Schmieder}, B., {Demoulin}, P., {Aulanier}, G., \& {Golub}, L. 1996, \apj, 467, 881

\bibitem[{{Sen} \& {Moreno-Insertis}(2025)}]{Sen-FMI:2025}
{Sen}, S. \& {Moreno-Insertis}, F. 2025, \aap, 699, A106

\bibitem[{{Sen} {et~al.}(2024){Sen}, {Prasad}, {Liakh}, \& {Keppens}}]{Sen:2024}
{Sen}, S., {Prasad}, A., {Liakh}, V., \& {Keppens}, R. 2024, \aap, 688, A64

\bibitem[{{Shibata} {et~al.}(1992){Shibata}, {Ishido}, {Acton}, {Strong}, {Hirayama}, {Uchida}, {McAllister}, {Matsumoto}, {Tsuneta}, {Shimizu}, {Hara}, {Sakurai}, {Ichimoto}, {Nishino}, \& {Ogawara}}]{1992PASJ...44L.173S}
{Shibata}, K., {Ishido}, Y., {Acton}, L.~W., {et~al.} 1992, \pasj, 44, L173

\bibitem[{{Shibata} {et~al.}(2007){Shibata}, {Nakamura}, {Matsumoto}, {Otsuji}, {Okamoto}, {Nishizuka}, {Kawate}, {Watanabe}, {Nagata}, {UeNo}, {Kitai}, {Nozawa}, {Tsuneta}, {Suematsu}, {Ichimoto}, {Shimizu}, {Katsukawa}, {Tarbell}, {Berger}, {Lites}, {Shine}, \& {Title}}]{2007Sci...318.1591S}
{Shibata}, K., {Nakamura}, T., {Matsumoto}, T., {et~al.} 2007, Science, 318, 1591

\bibitem[{{Sivaraman} {et~al.}(1992){Sivaraman}, {Rausaria}, \& {Aleem}}]{1992SoPh..138..353S}
{Sivaraman}, K.~R., {Rausaria}, R.~R., \& {Aleem}, S.~M. 1992, \solphys, 138, 353

\bibitem[{{Soenen} {et~al.}(2009){Soenen}, {Zuccarello}, {Jacobs}, {Poedts}, {Keppens}, \& {van der Holst}}]{2009:soenen}
{Soenen}, A., {Zuccarello}, F.~P., {Jacobs}, C., {et~al.} 2009, \aap, 501, 1123

\bibitem[{{Sterling} {et~al.}(2017){Sterling}, {Moore}, {Falconer}, {Panesar}, \& {Martinez}}]{2017ApJ...844...28S}
{Sterling}, A.~C., {Moore}, R.~L., {Falconer}, D.~A., {Panesar}, N.~K., \& {Martinez}, F. 2017, \apj, 844, 28

\bibitem[{{T{\"o}r{\"o}k} {et~al.}(2004){T{\"o}r{\"o}k}, {Kliem}, \& {Titov}}]{2004A&A...413L..27T}
{T{\"o}r{\"o}k}, T., {Kliem}, B., \& {Titov}, V.~S. 2004, \aap, 413, L27

\bibitem[{{Xia} {et~al.}(2018){Xia}, {Teunissen}, {El Mellah}, {Chan{\'e}}, \& {Keppens}}]{2018ApJS..234...30X}
{Xia}, C., {Teunissen}, J., {El Mellah}, I., {Chan{\'e}}, E., \& {Keppens}, R. 2018, \apjs, 234, 30

\bibitem[{{Yan} {et~al.}(2018){Yan}, {Yang}, {Xue}, {Mei}, {Kong}, {Wang}, \& {Li}}]{Yan:2018}
{Yan}, X.~L., {Yang}, L.~H., {Xue}, Z.~K., {et~al.} 2018, \apjl, 853, L18

\bibitem[{{Yeates}(2024)}]{Yeates:2024}
{Yeates}, A.~R. 2024, \solphys, 299, 83

\bibitem[{{Young} \& {Muglach}(2014)}]{2014PASJ...66S..12Y}
{Young}, P.~R. \& {Muglach}, K. 2014, \pasj, 66, S12

\bibitem[{{Zhang} {et~al.}(2021){Zhang}, {Liu}, {Wang}, {Zhou}, {Zhuang}, \& {Li}}]{Zhang:2021}
{Zhang}, Q., {Liu}, R., {Wang}, Y., {et~al.} 2021, \aap, 647, A171

\bibitem[{{Zhao} {et~al.}(2015){Zhao}, {DeVore}, {Antiochos}, \& {Zurbuchen}}]{Zhao:2015}
{Zhao}, L., {DeVore}, C.~R., {Antiochos}, S.~K., \& {Zurbuchen}, T.~H. 2015, \apj, 805, 61

\bibitem[{{Zhao} {et~al.}(2017){Zhao}, {Xia}, {Keppens}, \& {Gan}}]{zhao2017}
{Zhao}, X., {Xia}, C., {Keppens}, R., \& {Gan}, W. 2017, \apj, 841, 106

\end{thebibliography}

\end{document}